\documentclass{aa}
\usepackage{graphics}
\usepackage{graphicx}
\usepackage{amssymb}
\begin{document}

\def\deg{$^{\rm o}$}
\def\ico{$I_{\rm CO}$}
\def\b14{$I_{\rm 1.4}$}
\def\Qco{$Q_{\rm CO/RC}$}
\def\qco{$q_{\rm CO/RC}$}
\def\qfir{$q_{\rm FIR/RC}$}

\title{The Molecular Connection to the FIR-Radio Continuum 
Correlation in Galaxies}

\author{M. Murgia \inst{1,2} \and T.T. Helfer\inst{3} \and R. Ekers\inst{4} 
\and L. Blitz\inst{3} 
\and L. Moscadelli\inst{2} \and T. Wong\inst{4,5} \and R. Paladino\inst{2,6}}

\institute{
Istituto di Radioastronomia del CNR, Via Gobetti 101, I-40129 Bologna, Italy
\and
INAF\,-\,Osservatorio Astronomico di Cagliari, Loc. Poggio dei Pini, Strada 54,
I-09012 Capoterra (CA), Italy
\and
Radio Astronomy Laboratory, University of California, Berkeley, CA 94720, USA
\and
Australia Telescope National Facility, CSIRO, P.O. Box 76, Epping, N.S.W., 1710, Australia
\and
School of Physics, University of New South Wales, Sydney, NSW 2052, Australia
\and
Dipartimento di Fisica, Universit{\`a} di Cagliari, Cittadella Universitaria, I-09042 Monserrato (CA), Italy
}

\date{Received; Accepted}

\abstract{We have studied the relationships between the radio continuum (RC) 
and CO emission for a set of galaxies selected from the BIMA Survey of
Nearby Galaxies. 
We find that the global CO-RC correlation is as tight as the global
FIR-RC correlation for the 24 galaxies studied.  
Within 9 galaxies with $\sim$6\arcsec\
CO and RC data available, the CO and RC emission is as
tightly correlated as its global value; the radially averaged 
correlation is nearly linear, extends over four order of magnitude and holds down to
 the smallest linear resolution of the observations, which is $\sim100$ pc. 
We define \qco\ as the log of the ratio of the CO to RC flux as
a way to characterize the CO-RC correlation.  
Combining 6\arcsec\ pixel-by-pixel comparisons across all sources yields an 
average small-scale correlation of \qco\ = 1.1 $\pm$ 0.28; that is,
the spatially resolved correlation has a dispersion that
is less than a factor of 2.
There are however systematic variations in the CO/RC ratio; the strongest organized
structures in \qco\ tend to be found along spiral arms and on size 
scales much larger than the resolution of the observations.
We do not measure any systematic trend in CO/RC ratio as a function of
radius in galaxies.  The constancy of the CO/RC ratio stands
in contrast to the previously measured decrease in the FIR/RC ratio as a 
function of radius in galaxies. 
We suggest that the excellent correlation between the CO, RC and
FIR emission in galaxies is a consequence of regulation by
hydrostatic pressure; this model links all three
emissions without invoking an explicit dependence on a star
formation scenario.
\keywords{radio continuum: galaxies - galaxies:spiral - ISM: molecules - stars: formation
}}

\offprints{M. Murgia, m.murgia@ira.cnr.it}

\titlerunning{The Molecular Connection to the FIR-Radio Continuum Correlation in Galaxies}
\authorrunning{M. Murgia et al.}
\maketitle

\section{Introduction}

The correlation between the global far-infrared (FIR) and
centimeter-wavelength radio continuum (RC) emissions in galaxies 
is at the same time one of the most robust and one of the most puzzling 
relations in extragalactic research.  
On global scales, the FIR and RC emissions are
linearly correlated over five orders  of magnitude of
luminosity, with an RMS scatter of less than a factor of 2
(Condon 1992; Yun, Reddy, \& Condon 2001).  The relation
holds across a remarkable variety of galaxies to include dwarf 
galaxies (Leroy et al. 2004), normal barred and unbarred spiral galaxies 
of all Hubble types, irregular galaxies, E/S0 galaxies with ongoing 
star formation, starbursts, Seyferts and radio-quiet quasars (Condon 
1992 and references therein), out to the most extreme starforming galaxies 
in the early Universe, at $z$ = 1 and beyond (Appleton et al. 2004).

Although both the FIR and RC emission are thought to originate as
a result of star formation in molecular clouds, the emission
mechanisms for the two processes are completely different.
The FIR arises from dust heated by
newly-born massive stars, still embedded in their parent envelopes.
The RC is, for the most part, nonthermal synchrotron emission that arises 
from the interaction of relativistic electrons with the ambient 
magnetic field in which they diffuse. 
The relativistic electrons are thought to be 
accelerated in supernova remnants, which result from the
massive stars.  
Given the many disparate steps, the completely different physics,
and the different time scales for the various processes
involved, the tightness of the global FIR-RC correlation is
surprising indeed.

What is known about the relationship of molecular gas to the FIR
and RC distributions?
The FIR emission is well correlated with the
CO emission on large scales in galaxies (e.g. Devereux \& 
Young 1990; Young \& Scoville 1991 and references therein), 
and the CO emission is also well correlated with the RC
on global scales (Rickard, Turner \& Palmer 1977;
Israel \& Rowan-Robinson 1984; 
Adler, Allen \& Lo 1991; Murgia et al. 2002).  
At intermediate resolutions of $\sim$1--3 kpc, Marsh \&
Helou (1995) found a small but systematic decrease in the
FIR/RC ratio as a function of increasing radial
distance, which they interpreted as a smearing of the RC by the
propagation of the radiating electrons.  By contrast,
Adler et al. (1991) and
Murgia et al. (2002) found that the ratio between the RC and
the CO emission is constant, to within a factor of 3, both inside the same
galaxy and from galaxy to galaxy, down to kpc size scales.

The arcminute-scale resolution of IRAS and ISO has limited
extragalactic comparisons of the FIR-RC correlation to
kpc sizes for all but a few Local Group sources.  However,
on size scales of a few hundred parsecs and smaller, detailed maps 
of gas within several kpc of the Sun suggest that the radio continuum
emission associated with regions of star formation is,
for the most part, thermal, and does not follow the FIR-RC
correlation (Boulanger \& Perault 1988;
Haslam \& Osbourne 1987; Wells 1997).  
Why there should be a good global
correlation between the CO, FIR and the synchrotron-dominated RC
is thus not at all expected from observations of the local
interstellar medium.  
Furthermore, given the good correlations
on large scales in galaxies, and the poor local correlation,
one might expect that there may be some linear size scale(s)
at which the CO-RC and FIR-RC correlations break down.

In this paper, we examine several aspects of the CO-RC-FIR correlation
problem.  First, we quantify the FIR-RC, CO-RC and CO-FIR
correlations on global scales for a sample of 24 galaxies with
improved integrated single-dish CO measurements.  Then, using 6\arcsec\
imaging for a subsample of 9 sources, we show that the CO-RC relation 
is equally tight {\it within} galaxies as it is from galaxy to galaxy, 
down to linear scales of $\sim$100 pc.  
We show that the most significant large-scale variations in the
CO-RC relation appear to be associated with spiral arms and that
there is not a significant radial variation in the ratio.
Finally, we propose that the RC-FIR-CO correlations may be
explained on large and small scales to be a result of 
regulation by the hydrostatic pressure in galaxies.

\section{Sample Selection and Data Reduction}

Galaxies were selected from the Berkeley-Illinois-Maryland 
Association Survey of Nearby Galaxies (BIMA SONG), an imaging survey of 
the CO (J=1-0) emission in 44 nearby spiral galaxies with a typical 
resolution of 6\arcsec~(Regan et al. 2001; Helfer et al. 2003). 
For 24 of the galaxies in the BIMA SONG sample, 
fully-sampled single-dish CO images were incorporated into the BIMA SONG
images, so that all of the CO flux is imaged for these galaxies
(i.e., there is no ``missing flux'' problem for these sources).
The single-dish data for these 24 sources, taken at 55\arcsec\ resolution 
with the NRAO 12m telescope\footnote{The National Radio Astronomy 
Observatory is a facility of the National Science Foundation operated 
under cooperative agreement by Associated Universities, Inc.},
comprise the most extensive collection of
fully-sampled, two-dimensional single-dish CO images of external
galaxies to date, and we use global CO flux measurements from these 
single-dish data.  

From the BIMA SONG data set, we selected a sub-sample of nine 
CO-bright galaxies for which high quality VLA B-array at 1.4 GHz were 
available either from the NRAO VLA archive$^{1}$ or from 
the Faint Images of the Radio Sky at Twenty-Centimeters (FIRST) 
survey (White et al. 1997).
The VLA archival data were calibrated and imaged with the NRAO package 
AIPS following standard procedures. To improve the sensitivity of the 
VLA radio continuum images to large scale structures, D array data
were combined with the B array data. The D array data were taken
from the VLA archive when available, and combined in the visibility
plane with the B array data; otherwise D array images from the 
NRAO VLA Sky Survey (NVSS, Condon et al. 1998) were added in the 
image plane using the linear combination method implemented in the 
MIRIAD task IMMERGE (Sault, Teuben, \& Wright 1995).

The subsample of 9 galaxies considered in this work are listed in 
Table~\ref{bimasample} 
along with some of their physical parameters.
All of them have an image of the CO integrated intensity from the 
NRAO 12m telescope.
We therefore have CO and RC images at two different angular resolutions: 
a ``low'' resolution ($\sim55$\arcsec) pair obtained from the NRAO 12m and 
VLA D-array, and a ``high'' resolution ($\sim6$\arcsec) pair.
The RC and CO image characteristics for the low and high resolution data sets 
are listed in Table~\ref{imageslr} and \ref{imageshr}, respectively,
with beam sizes listed in angular and linear scales.

We emphasize that these data are among the most comprehensive
and uniformly processed CO and RC data from normal galaxies.
Furthermore, the angular resolution of the CO and RC images are
well matched, not just by filtering the data, but by having comparable
intrinsic coverage over the relevant spatial frequencies.
Both the low-resolution and high-resolution data sets had
intrinsic beam areas where the RC and CO images matched typically to
within 10\%.  Nonetheless, to account for even small differences 
in the beam areas, we convolved the images to common angular
resolution in the quantitative analysis below.

Finally, in order to study the RC, CO and FIR global correlations, the 
integrated 60 $\mu$m fluxes of the galaxies in our sample have been 
taken from the IRAS and IRAS HIRES ($\ga1$\arcmin) databases (see
Table~\ref{globalfluxes} for references). 

\begin{table*}
\caption{Galaxy Sample}
\begin{center}
\begin{tabular}{lllllll}
\noalign{\smallskip}
\hline                  
\noalign{\smallskip}    
Galaxy   & $\alpha$(J2000) & $\delta$(J2000) & d & i & P.A. & RC3 Type, Nucleus  \\
         & ($h$ $m$ $s$)   & (\degr\ \arcmin\ \arcsec) & (Mpc)  & (\deg)  & (\deg) &                  \\ 
\hline
IC 342   &  03 46 49.7  & 68 05 45 & 3.9   & 31 &  37  & SAB(rs)cd,  HII  \\
NGC 4258 &  12 18 57.5  & 47 18 14 & 8.1   & 65 &  176 & SAB(s)bc, Sy1.9  \\
NGC 4414 &  12 26 27.2  & 31 13 24 & 19.1  & 55 &  159 & SA(rs)c?, T2:    \\
NGC 4736 &  12 50 53.06 & 41 07 14 & 4.3   & 35 &  100 & (R)SA(r)ab,T2    \\
NGC 5005 &  13 10 56.23 & 37 03 33 & 21.3  & 61 &  65  & SAB(rs)bc, L1.9  \\
NGC 5033 &  13 13 27.53 & 36 35 38 & 18.7  & 62 &  170 & SA(s)c, Sy1.5    \\
NGC 5055 &  13 15 49.25 & 42 01 49 & 7.2   & 55 &  105 & SA(rs)bc, T2     \\
NGC 5194 &  13 29 52.35 & 47 11 54 & 7.7   & 15 &  0   & SA(s)bc pec, Sy2 \\
NGC 6946 &  20 34 52.33 & 60 09 14 & 5.5   & 54 &  65  & SAB(rs)cd,HII    \\
\hline
\noalign{\smallskip}
\label{bimasample}
\end{tabular}
\end{center}
\begin{list}{}{}
\item[] 
Coordinates, distances, inclinations, position angles, RC3 type and nuclear 
classifications from Helfer et al. (2003).
\end{list}
\end{table*}

\begin{table*}
\caption{Low resolution CO and RC image characteristics.}
\begin{center}
\begin{tabular}{lllllll}
\noalign{\smallskip}
\hline                  
\noalign{\smallskip}    
Galaxy   & VLA data      &   RC-$\theta_{maj}\times\theta_{min}$  &  $\sigma_{RC}$  &  CO-$\theta_{maj}\times\theta_{min}$ & $D_{maj}$ $\times$ $D_{maj}^{~a)}$ & $\sigma_{CO}^{~b)}$      \\
         &               &   (\arcsec$\times$\arcsec)                 &     (mJy~bm$^{-1}$)     &    (\arcsec$\times$\arcsec)    & (kpc $\times$ kpc)    &   (Jy~bm$^{-1}$~km~s$^{-1}$)  \\ \hline
IC 342   & D & 52$\times$44 & 0.15 & 55$\times$55 & 1.0 $\times$ 1.0 & 23.3 \\  
NGC 4258 & D & 45$\times$45 & 0.50 & 55$\times$55 & 2.1 $\times$ 2.1 & 31.4 \\
NGC 4414 & C & 14$\times$14 & 0.20 & 55$\times$55 & 5.0 $\times$ 5.0 & 36.2 \\
NGC 4736 & D & 54$\times$54 & 0.13 & 55$\times$55 & 1.1 $\times$ 1.1 & 18.4 \\
NGC 5005 & D & 48$\times$48 & 0.20 & 55$\times$55 & 5.6 $\times$ 5.6 & 54.7 \\
NGC 5033 & D & 54$\times$54 & 0.20 & 55$\times$55 & 4.9 $\times$ 4.9 & 44.8 \\
NGC 5055 & D & 54$\times$54 & 0.20 & 55$\times$55 & 1.9 $\times$ 1.9 & 35.0 \\
NGC 5194 & D & 54$\times$54 & 0.30 & 55$\times$55 & 2.0 $\times$ 2.0 & 24.7 \\
NGC 6946 & D & 44$\times$40 & 0.10 & 55$\times$55 & 1.5 $\times$ 1.5 & 35.0 \\ 
\hline
\noalign{\smallskip}
\label{imageslr}
\end{tabular}

\end{center}
\begin{list}{}{}
\item[$a)$] The linear resolution refers to the CO beam.
\item[$b)$] $\sigma_{CO}$ is the rms level in the unclipped image of 
integrated intensity.
\end{list}
\end{table*}

\begin{table*}
\caption{High resolution CO and RC image characteristics.}
\begin{center}
\begin{tabular}{lllllll}
\noalign{\smallskip}
\hline                  
\noalign{\smallskip}    
Galaxy   & VLA data      &   RC-$\theta_{maj}\times\theta_{min}$  &  $\sigma_{RC}$  &  CO-$\theta_{maj}\times\theta_{min}$ & $D_{maj}\times D_{min}^{~a)}$ & $\sigma_{CO}^{~b)}$      \\
         &               &   (\arcsec$\times$\arcsec)                 &     (mJy~bm$^{-1}$)     &    (\arcsec$\times$\arcsec)    & (pc$\times$pc)    &   (Jy~bm$^{-1}$~km~s$^{-1}$)  \\ \hline
IC 342   & B+D & 5.5$\times$5.0 & 0.05 & 5.6$\times$5.1 & 110$\times$97 & 4.8 \\  
NGC 4258 & AnB+D & 3.5$\times$3.3 & 0.05 & 6.1$\times$5.4 & 240$\times$210 & 4.8 \\
NGC 4414 & B+D & 5.4$\times$5.4 & 0.08 & 6.4$\times$5.0 & 590$\times$460 & 1.9 \\
NGC 4736 & B+D & 6.1$\times$5.1 & 0.05 & 6.9$\times$5.0 & 140$\times$100 & 3.5 \\
NGC 5005 & B+D & 5.4$\times$5.0 & 0.03 & 6.2$\times$6.0 & 640$\times$620 & 2.9 \\
NGC 5033 & B+D & 5.3$\times$5.0 & 0.07 & 6.1$\times$5.4 & 550$\times$490 & 5.6 \\
NGC 5055 & B+D & 5.5$\times$5.0 & 0.04 & 5.8$\times$5.5 & 200$\times$190 & 3.6 \\
NGC 5194 & B+C & 6.0$\times$4.8 & 0.02 & 5.8$\times$5.1 & 214$\times$188 & 3.1 \\
NGC 6946 & B+C & 5.6$\times$4.9 & 0.10 & 6.0$\times$4.9 & 160$\times$130 & 3.1 \\
\hline
\noalign{\smallskip}
\label{imageshr}
\end{tabular}

\end{center}
\begin{list}{}{}
\item[$a)$] The linear resolution refers to the CO beam.
\item[$b)$] $\sigma_{CO}$ is the rms level in the unclipped image of 
integrated intensity.
\end{list}
\end{table*}

\section{Data Analysis}

In order to compare the point-to-point RC and CO brightnesses
across entire galaxy disks, we overlaid regular grids of 
rectangular beam-sized boxes on both the RC and CO images. 
We then calculated the 1.4 GHz RC brightness (\b14) and the CO 
integrated intensity (\ico) by averaging all pixel values within each box. 
For the 55\arcsec\ resolution data, we used CO images of integrated 
intensity made without clipping any of the channel images.  For
the 6\arcsec\ resolution images, we used CO images as presented in Helfer
et al. (2003), made by applying a smooth-and-mask technique that 
is very effective at showing low-level emission (but that may
introduce a bias against faint, compact emission; see van Gorkom
\& Ekers 1989; Helfer et
al. 2003).  The smooth-and-mask technique does not bias the
noise statistics of the final image, and we used the mask to
calculate the pixel-by-pixel rms in the high-resolution images.
For the radial profiles (below), the signal to noise ratio of the 
high- and low-resolution CO data is sufficient to use the unclipped 
images of the integrated intensity, avoiding any bias.

Fig.~1 shows the data analysis performed for the nine galaxies. 
For each galaxy, the upper and bottom left panels show the overlay of 
\b14~(contours) on \ico~(color) images respectively at low and 
high resolution. On top of the two panels, the intensity scale for the 
CO images is shown as a wedge. The RC contours start from a level of 5$\sigma$ 
and scale by a factor of $\sqrt{2}$.  

The upper right panels plot \b14~versus \ico~ at low (solid 
dots) and high (open dots) resolution.  In the bottom right corner 
insets, the upper and lower panels show the box grids for the low 
and high resolution images, respectively. Only points above 2$\sigma$ 
have been plotted.  Areas with obvious background continuum sources
were omitted from the analysis.

Finally, the bottom right panel shows the radial profiles of 
\b14~and \ico~(on the top) and their ratio \Qco $\equiv$ 
\ico(Jy sr$^{-1}$km s$^{-1}$)/\b14(mJy sr$^{-1}$) (on the bottom).  
For this analysis, we used
unclipped images of integrated intensity in CO to avoid
any bias against faint emission, and we calculated the mean 
intensities in concentric elliptical annuli
using the position angles and inclinations listed in 
Table~\ref{bimasample}.
We spaced the annuli to have widths along their major axis
at the resolution of the corresponding
images, or 6\arcsec\ for the high-resolution images and 55\arcsec\
for the low-resolution images. We plot only those points that
are measured at a level $\geq$1$\sigma$.
For each galaxy, the values of the mean and dispersion of \qco $\equiv$ 
$\log$(\Qco) are given on the bottom right plot of the radial variation
in \Qco.

\begin{figure*}[t]
\begin{center}
\includegraphics[width=16.5cm]{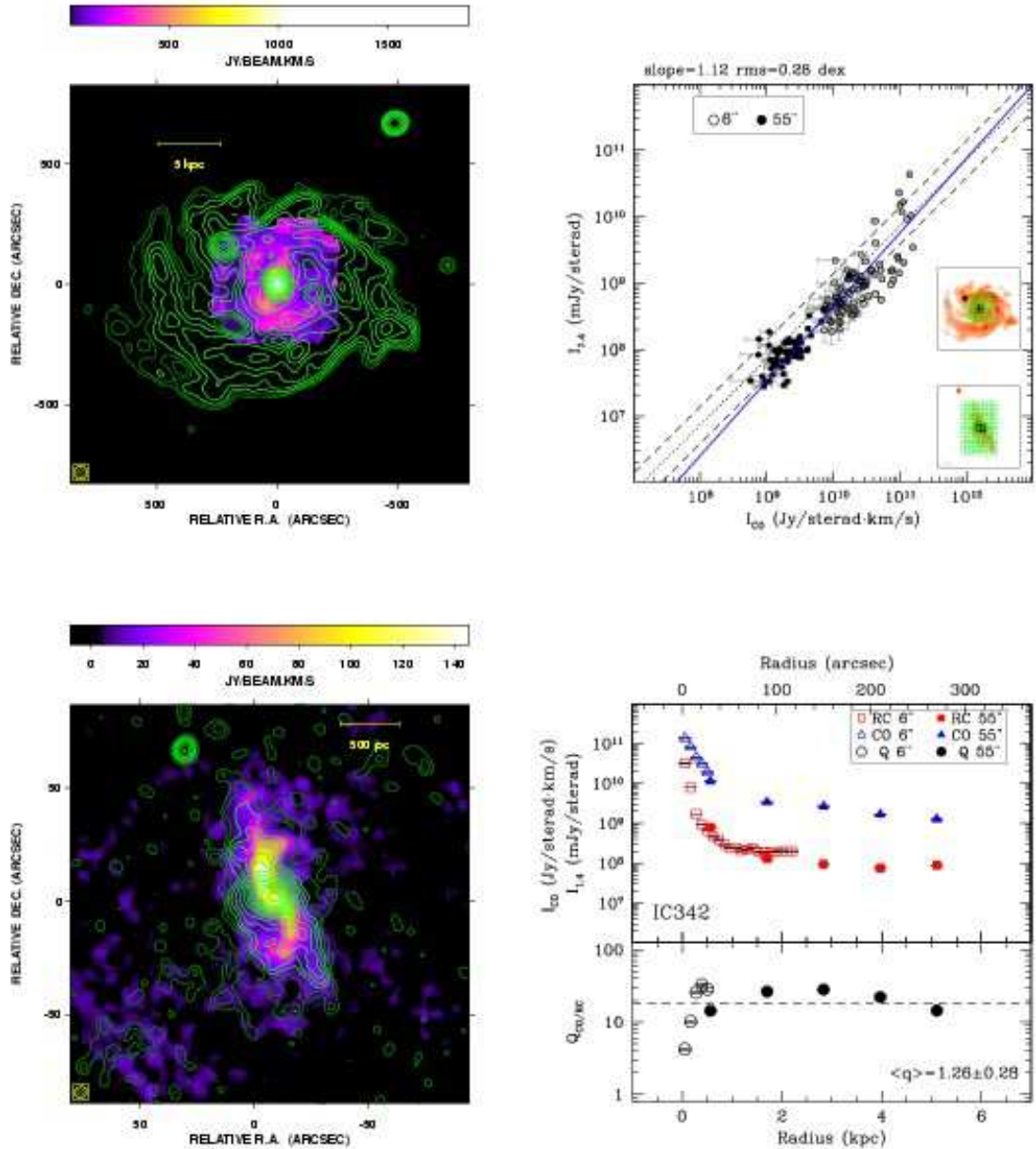}

\end{center}

\caption[] {Radio continuum and CO comparison in BIMA SONG galaxies.
{\it (Upper left)} RC contours overlaid on CO emission for 55\arcsec\ images.
{\it (Lower left)} RC contours overlaid on CO emission for 6\arcsec\ images.
Contours start at 5$\sigma$ and scale by a factor of $\sqrt{2}$; the 
bright contours are chosen to match the dynamic range of the CO emission.
{\it (Upper right)} \b14\ as a function of \ico\  for the 55\arcsec\ and
6\arcsec\ data sets.  The insets show the box grids used; the grid size
corresponds to the resolution, so that each point is essentially
independent.
The dotted (dashed) lines represent the mean (dispersion) of the 
RC/CO ratio for all nine sources.  The solid blue line is a weighted fit 
to the points shown which takes into account the errors in both coordinates.
{\it (Lower right)} Radial profiles of \ico\ and \b14\ {\it (top)} and
\Qco\ $\equiv$ \ico/\b14 {\it (bottom)} for the 55\arcsec\ and 6\arcsec\ images.
The mean and dispersion of \qco\ $\equiv \log (Q_{\rm CO/RC})$ is also given.
See ${\S}$3 for further details on the presentation; notes on
individual galaxies are given in Appendix A.  $(a)$ IC 342.}

\label{figura1}
\end{figure*}

\setcounter{figure}{0}

\begin{figure*}
\begin{center}
\includegraphics[width=16.5cm]{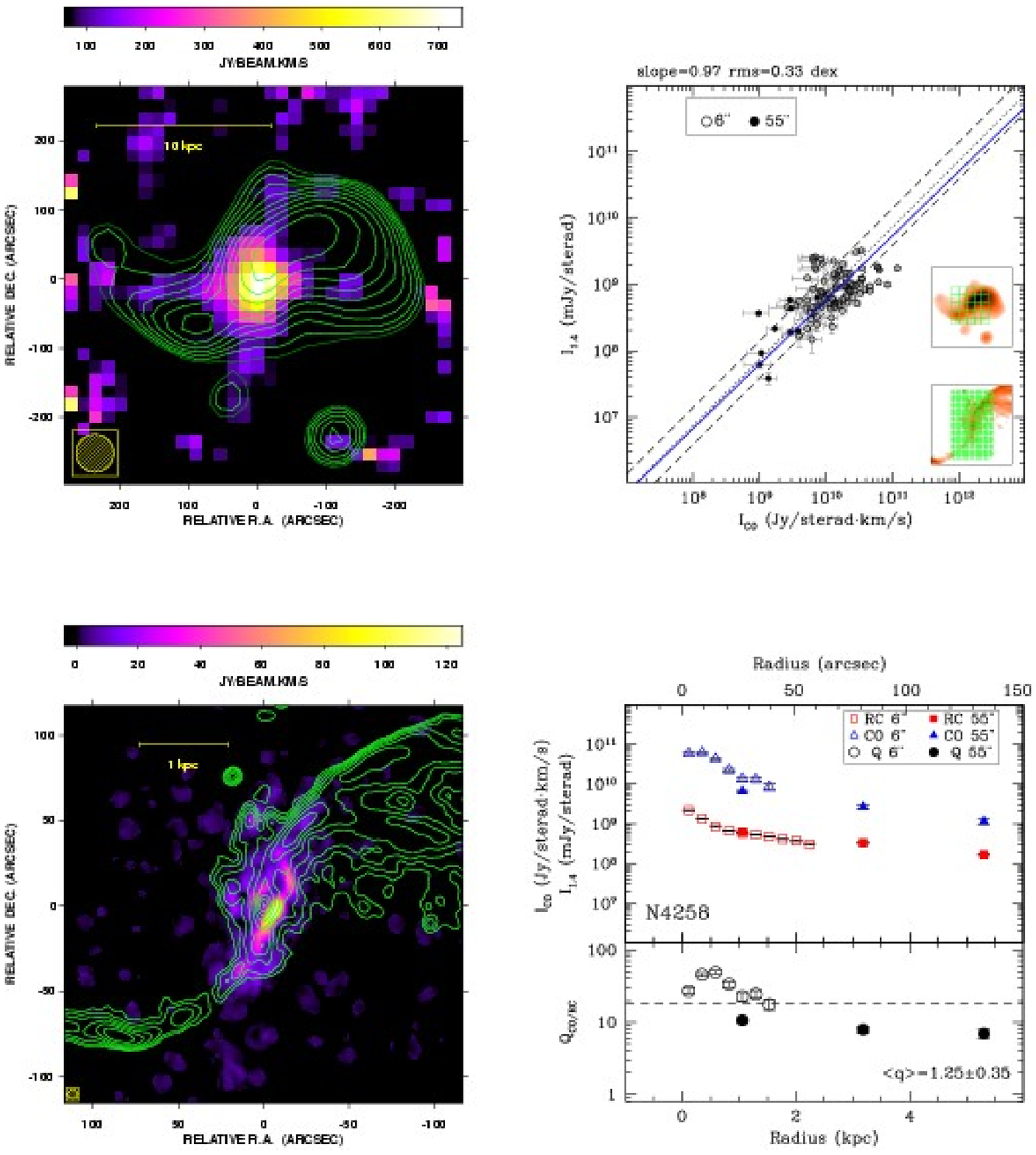}
\end{center}
\caption[] {$(b)$ NGC 4258. See Figure 1$a$.}
\end{figure*}

\setcounter{figure}{0}
\begin{figure*}
\begin{center}
\includegraphics[width=16.5cm]{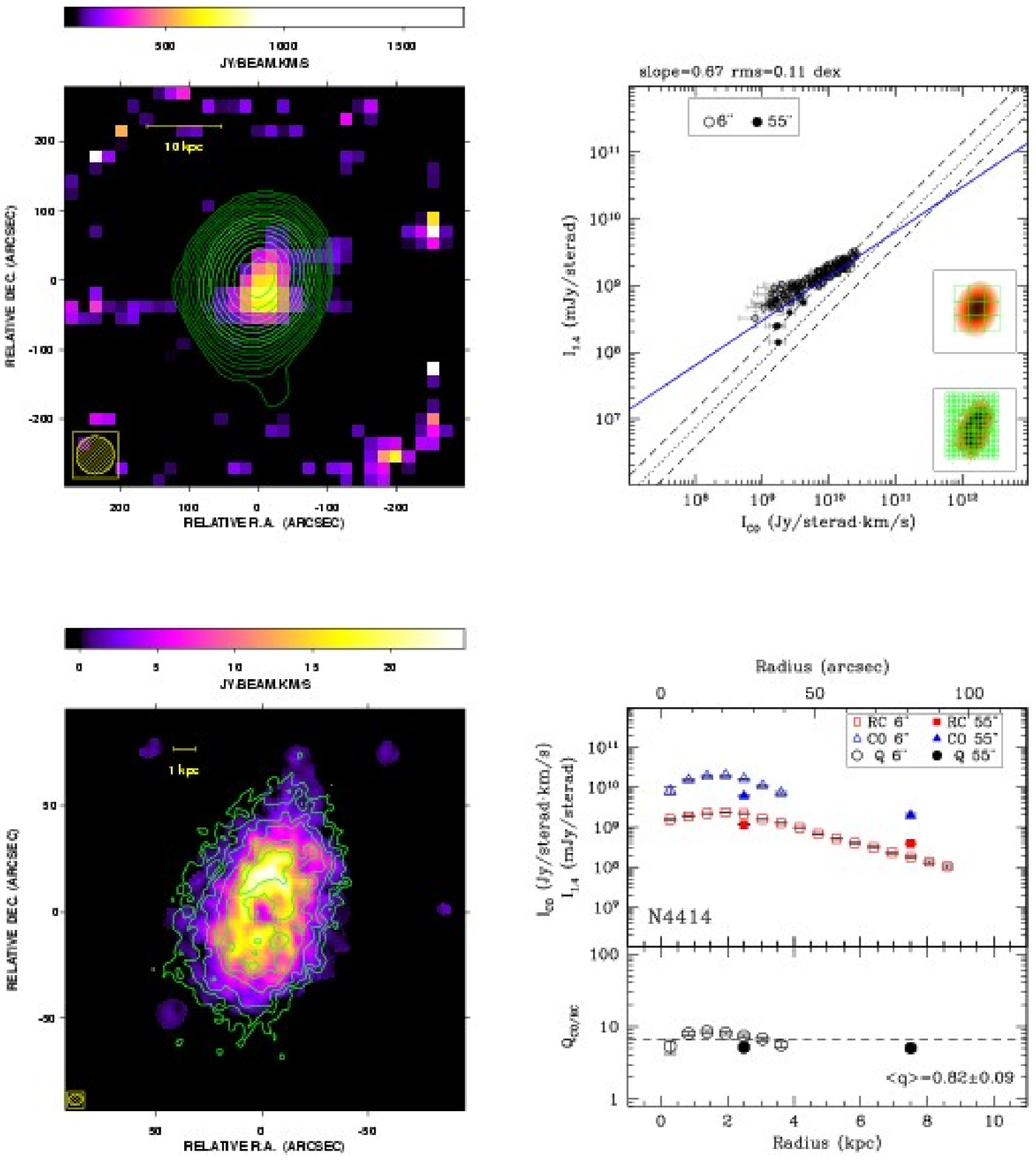}
\end{center}
\caption[] {$(c)$ NGC 4414. See Figure 1$a$.}
\end{figure*}

\setcounter{figure}{0}
\begin{figure*}
\begin{center}
\includegraphics[width=16.5cm]{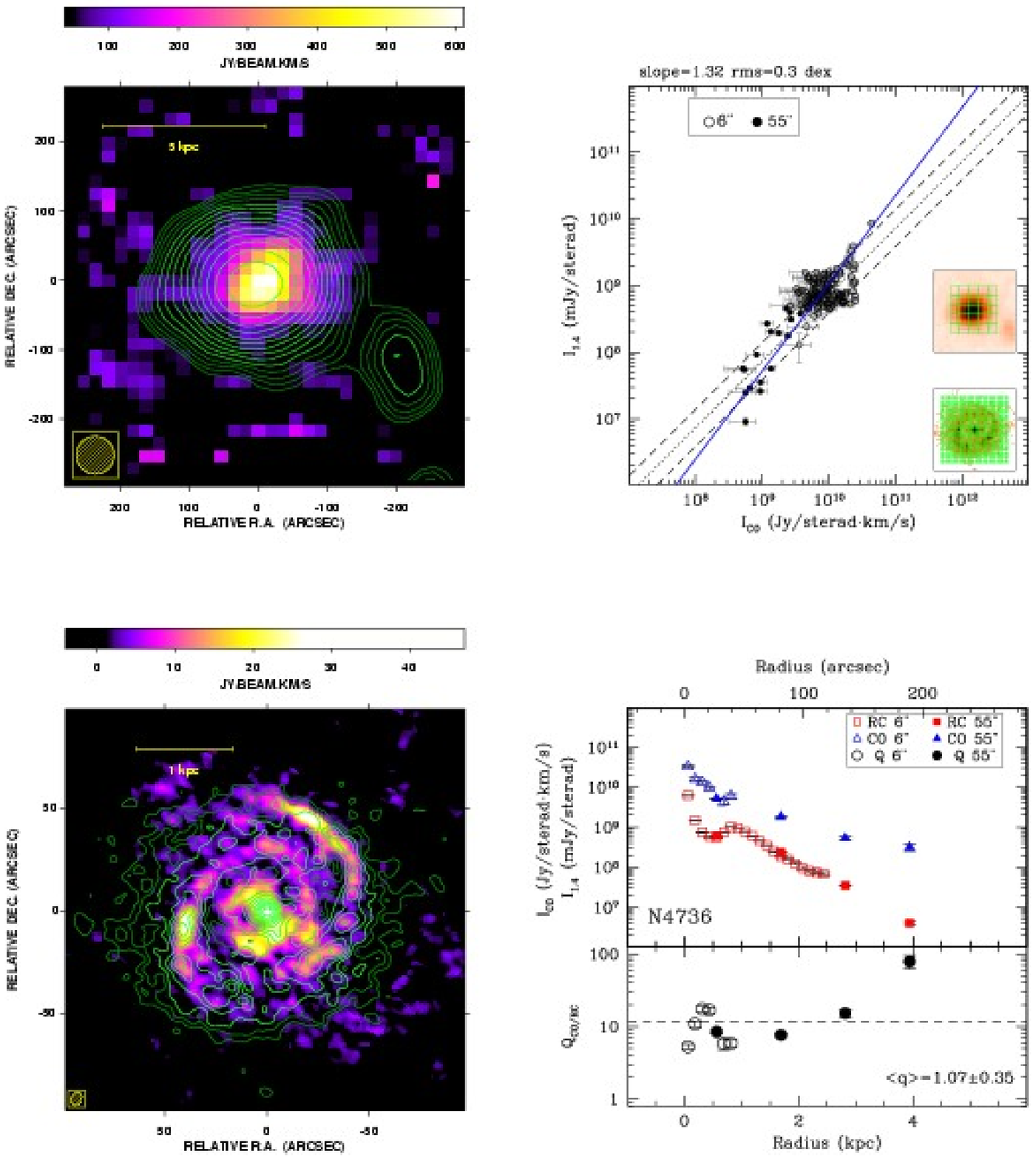}
\end{center}
\caption[] {$(d)$ NGC 4736. See Figure 1$a$.}
\end{figure*}

\setcounter{figure}{0}
\begin{figure*}
\begin{center}
\includegraphics[width=16.5cm]{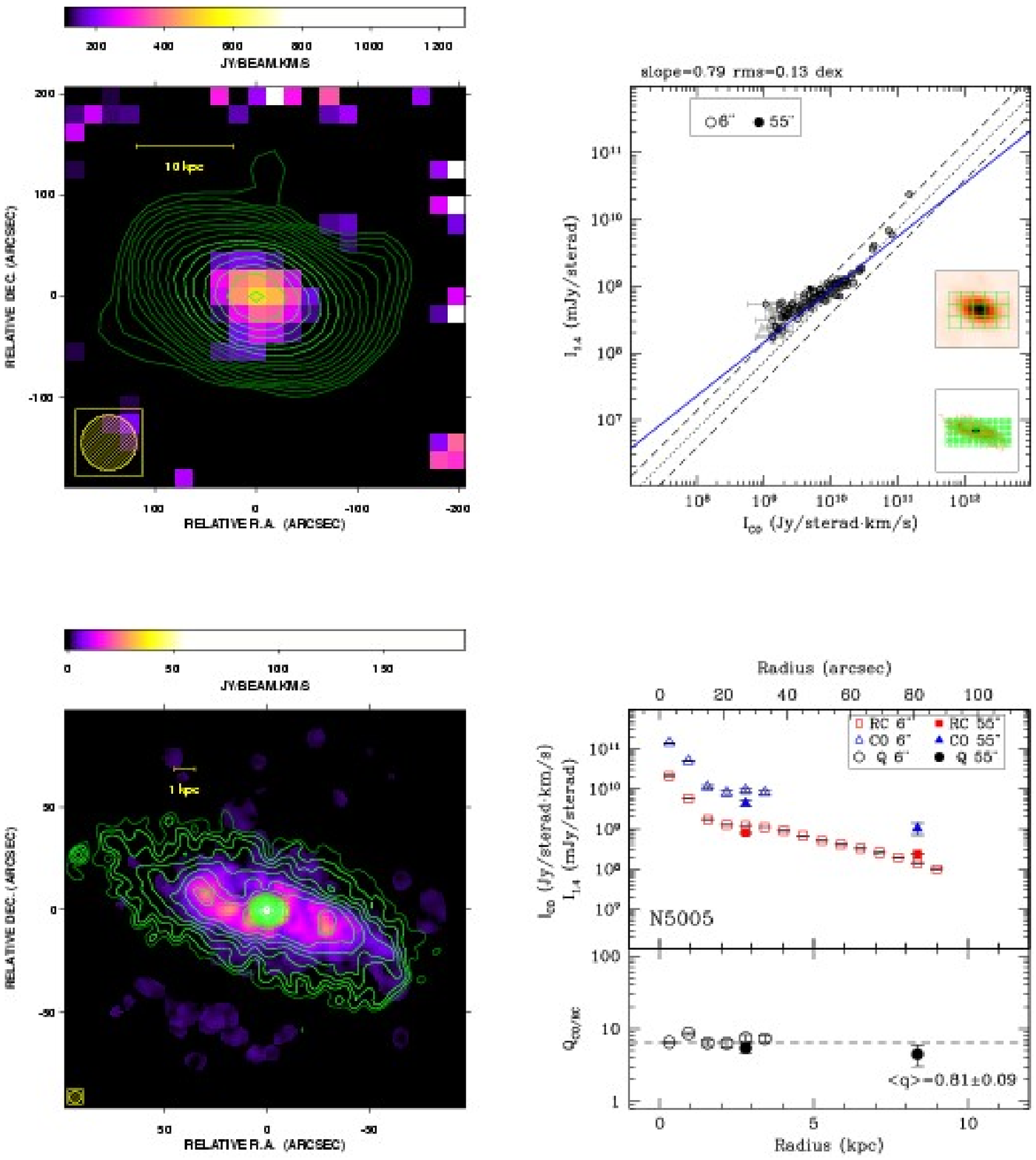}
\end{center}
\caption[] {$(e)$ NGC 5005. See Figure 1$a$.}
\end{figure*}

\setcounter{figure}{0}
\begin{figure*}
\begin{center}
\includegraphics[width=16.5cm]{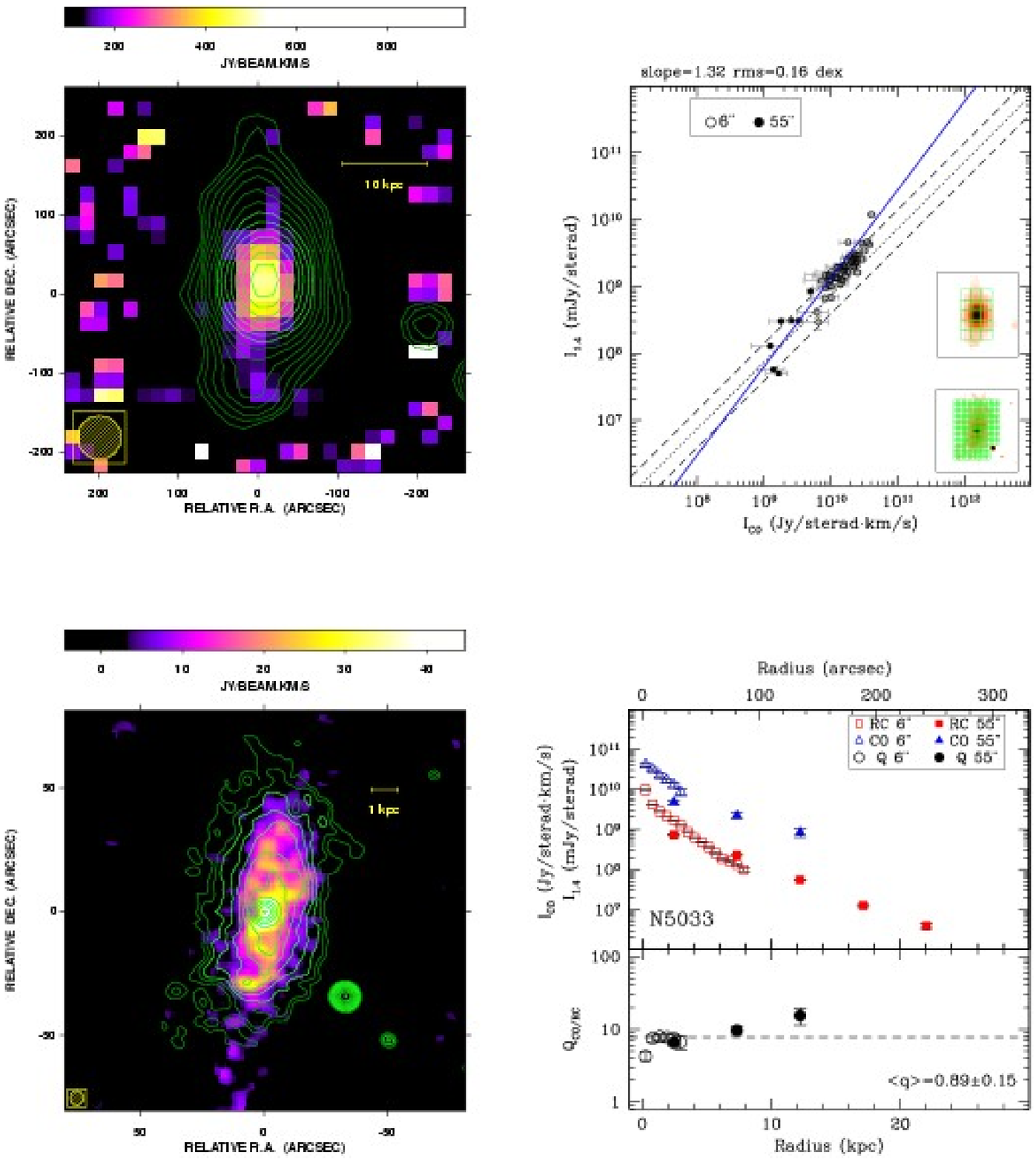}
\end{center}
\caption[] {$(f)$ NGC 5033. See Figure 1$a$.}
\end{figure*}

\setcounter{figure}{0}
\begin{figure*}
\begin{center}
\includegraphics[width=16.5cm]{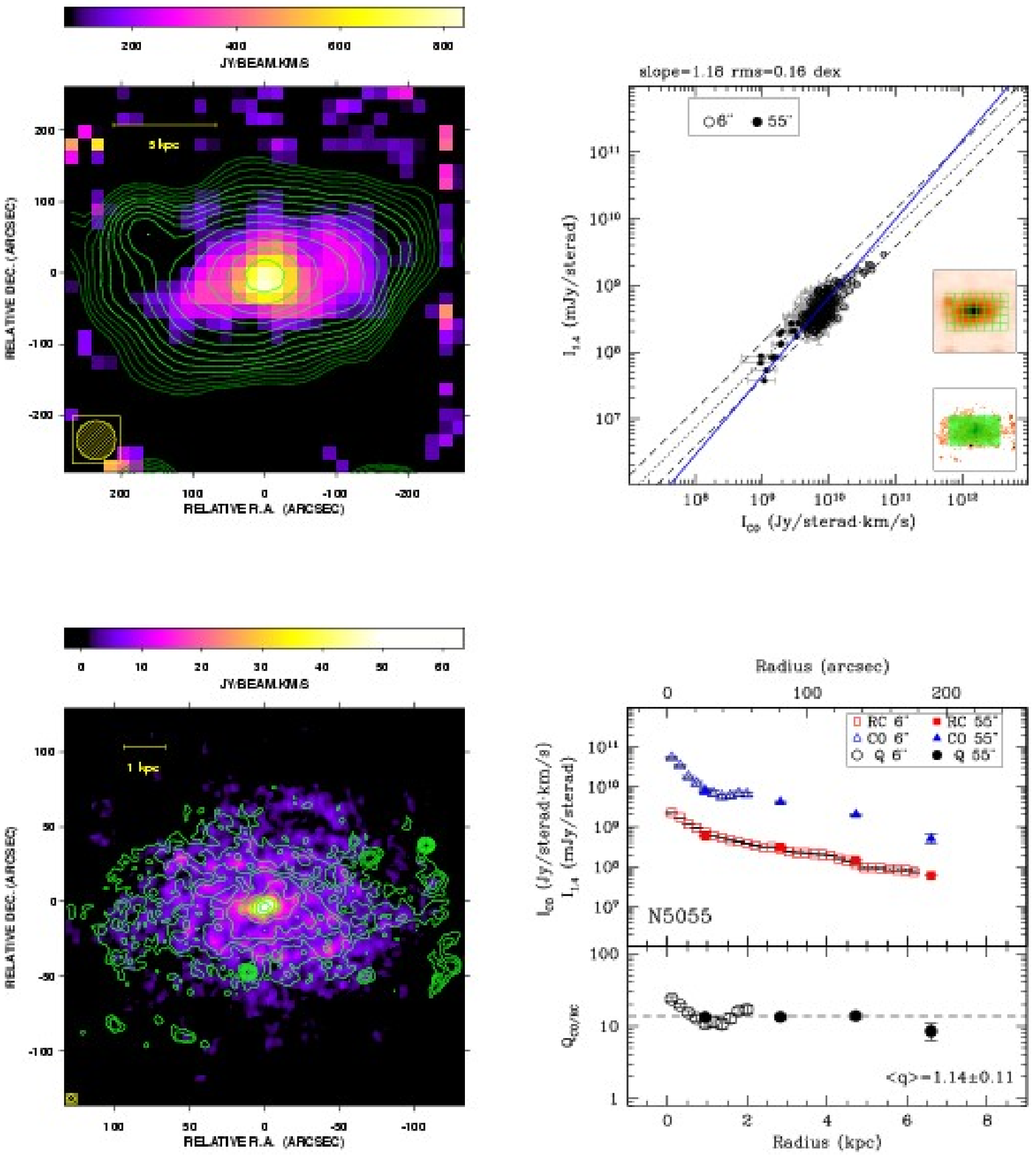}
\end{center}
\caption[] {$(g)$ NGC 5055. See Figure 1$a$.}
\end{figure*}

\setcounter{figure}{0}
\begin{figure*}
\begin{center}
\includegraphics[width=16.5cm]{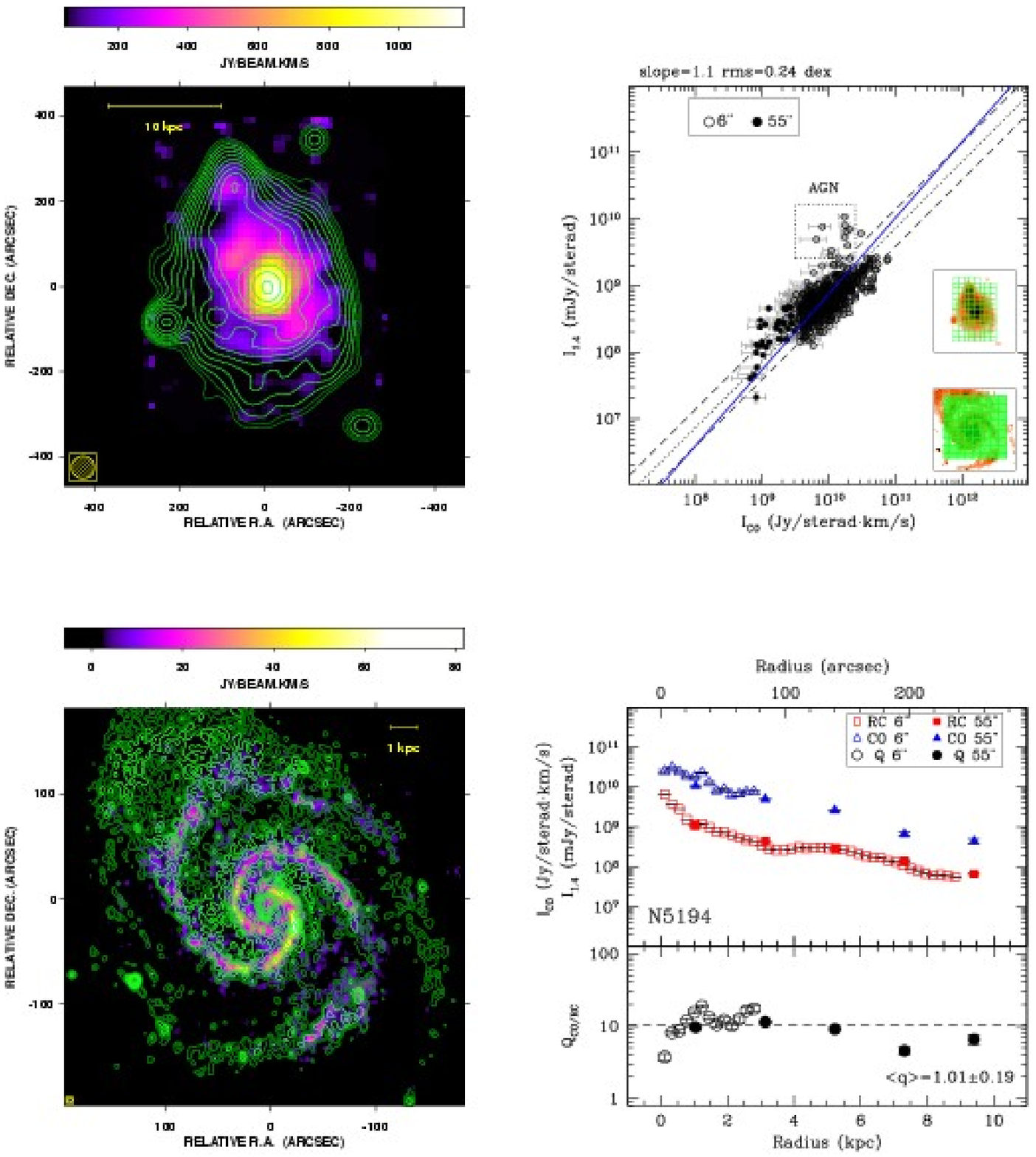}
\end{center}
\caption[] {$(h)$ NGC 5194. See Figure 1$a$.}
\end{figure*}

\setcounter{figure}{0}
\begin{figure*}
\begin{center}
\includegraphics[width=16.5cm]{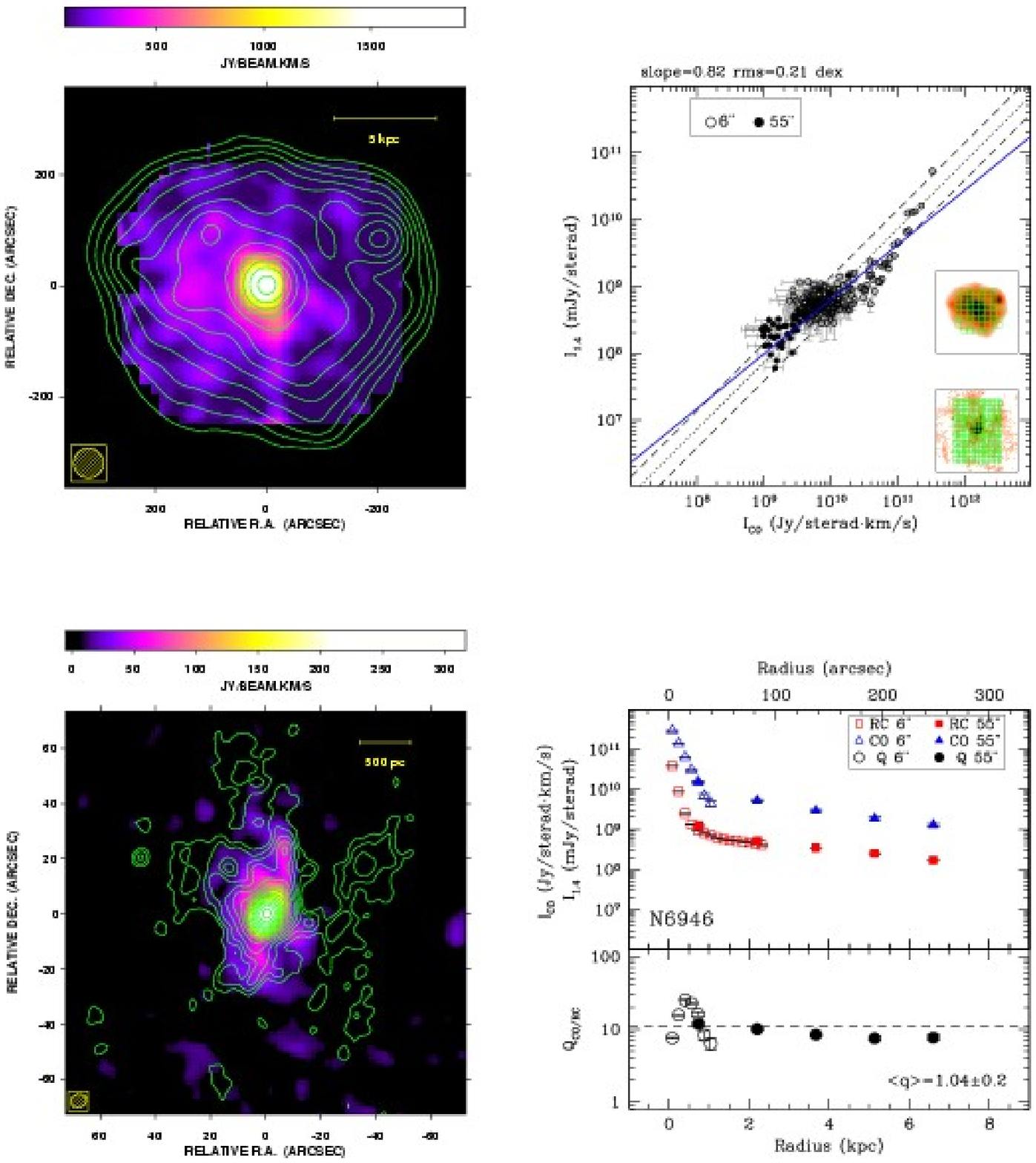}
\end{center}
\caption[] {$(i)$ NGC 6946. See Figure 1$a$.}
\end{figure*}

\begin{figure*}
\begin{center}
\includegraphics[height=6cm,angle=0]{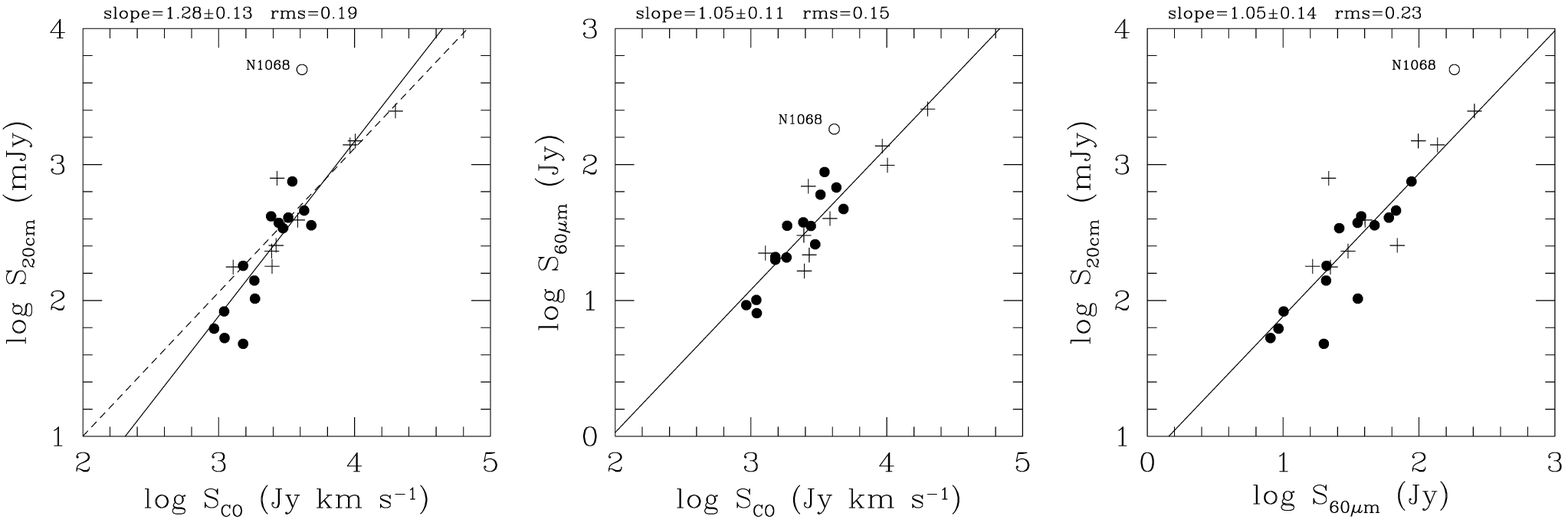}
\end{center}

\caption[] {Global CO-RC, FIR-CO, and RC-FIR fluxes for 24 BIMA SONG 
galaxies, shown with least squares fits (from which NGC 1068 is 
omitted).  The data are listed in Table~\ref{globalfluxes}.  Crosses 
are galaxies included in the high-resolution analysis of this paper. The dashed-line represents a least squares fit
 restricted to the data of the 9 galaxies analyzed in this work and yields a CO-RC correlation of $\log (S_{\rm RC}) \propto 1.06
 \pm 0.17 \log(S_{\rm CO})$.}
\label{figuraflux}
\end{figure*}

\section{Results}

\subsection{ Comparison of Global Fluxes}\label{sec:globflux}

For the 24 BIMA SONG galaxies that have associated single-dish CO
images, Helfer et al. (2003) tabulated CO global fluxes that were
directly measured from the single-dish images.
For these sources, we can quantify the global CO-RC, FIR-CO, and 
RC-FIR correlations.
The relevant fluxes are listed in Table~\ref{globalfluxes}, and the relations
are plotted in Fig.~\ref{figuraflux}.
Linear least squares
fits to the relations yield a CO-RC correlation of
log(S$_{1.4}$)/log(S$_{CO}$) $\propto$ 1.28$\pm$0.13,
a CO-FIR correlation 
log(S$_{FIR}$)/log(S$_{CO}$) $\propto$ 1.05$\pm$0.11,
and a FIR-RC correlation of
log(S$_{1.4}$)/log(S$_{FIR}$) $\propto$ 1.05$\pm$0.14.
(NGC 1068 is omitted from these fits, since it has a
significant contribution to its RC flux from its AGN;
however, omitting this source has a small impact on the
slopes derived.)
The RC-FIR correlation measured for these 23 sources is 
consistent with its best determination over a large (1809)
number of sources (Yun, Reddy \& Condon 2001), where the slope
of the RC/FIR fit (also using 1.4 GHz RC and 60 $\mu$m FIR)
is 0.99$\pm$0.01.

Of the three correlations measured here,
the slope of the CO-RC correlation is the steepest, and
is the only one that appears to be inconsistent with
a slope of unity.  
Furthermore, the CO-RC slope is consistent with that measured
by Murgia et al. (2002), or 1.3 $\pm$ 0.1, on arcminute
scales using independent data.
However, given the uncertainties, it is fair
to say that these data show that all three correlations 
are comparably tight, and that we cannot identify
any of the relations as being a ``fundamental'' one.

\subsection{Point-to-Point and Radial Correlations of the RC and CO 
Emissions}

The 55\arcsec\ RC and CO images are shown in the upper-left 
panels of Figure 1.
The RC emission, shown as contours in these figures, occasionally
includes point sources which are almost certainly background radio
sources.  Ignoring these background sources, we can
already see by eye that the general shape of the RC contours
follows the underlying CO emission reasonably well. There
are a few exceptions, like NGC 4258, for which the RC and CO 
emission seem to display quite different behaviours (see the Appendix 
for details).

The 6\arcsec\ resolution 
images are shown in the lower-left panels of Figure 1.  Here,
the detailed agreement is even more striking, partly because 
a larger dynamic range of emission is represented in these images.
We note that both the high- and low-resolution RC images have a 
larger dynamic range than the corresponding CO images; this prevents 
a full comparison of the two in the faintest regions and it may give
the impression that there are RC features that are unmatched by
the CO.  However, the opposite is not observed:  every CO feature
appears to have a RC counterpart.

\begin{figure*}
\begin{center}
\includegraphics[width=15cm]{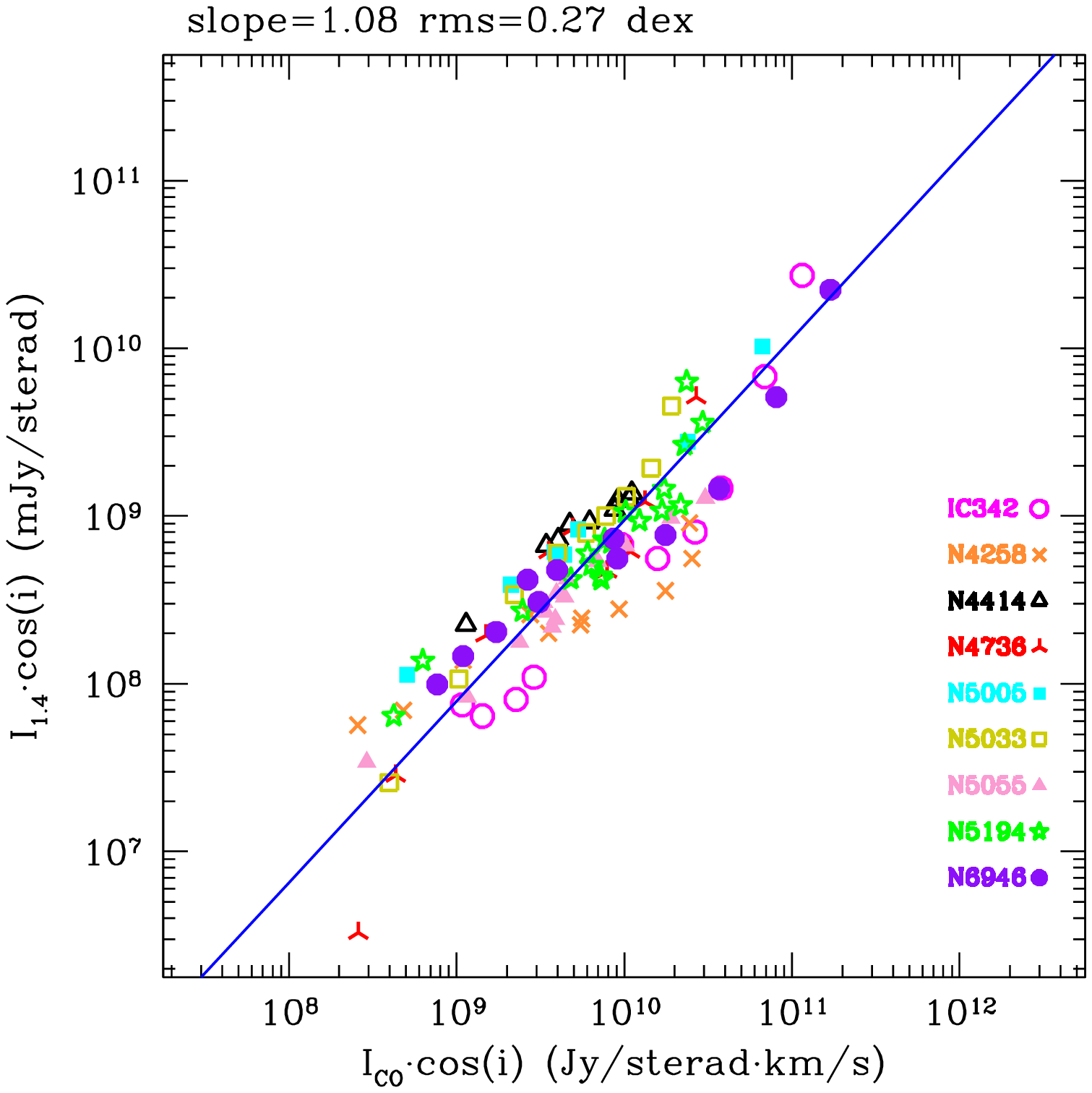}
\end{center}
\caption[] { CO-RC correlation in sample galaxies.  Each point represents
the average value in an annulus. The solid blue line is a weighted fit 
to the points shown which takes into account the errors in both coordinates and has a slope of 1.08.
}
\label{figuraradial}
\end{figure*}

The quantitative point-to-point comparison of the 
$I_{\rm 1.4}-I_{\rm CO}$ correlation, shown for both
the low and high resolution images in the upper-right
panels of Figure 1, shows an excellent agreement
between the two data sets for each galaxy.  First, we note that
for the point-to-point $I_{\rm 1.4}-I_{\rm CO}$ correlation, the
low-resolution data are aligned with the extrapolation of
the high-resolution data to lower intensity values. 
Moreover, the $I_{\rm 1.4}$ and $I_{\rm CO}$ radial profiles of the 
low and high resolution data also match very well (see bottom-right 
panels of Fig.\,1). This is particularly evident 
toward the central region of the galaxies where the two data sets 
overlap and is an effect that results from the incorporation 
of extended flux from the low-resolution images into the high-resolution
data.

The mean and dispersion in \qco\ are given in the bottom right
panels of Figure 1.  The dispersion within the individual sources 
ranges from 0.09 to 0.35; these correspond to dispersions in the CO/RC
flux ratios that are a factor of 1.2--2.2.  These are remarkably tight 
correlations given that the dynamic range extends over three
orders of magnitude for some sources.  The average \qco\ {\it
among} all nine sources is 1.03 $\pm$ 0.06 (uncertainty in the mean)
$\pm$ 0.17 (standard deviation); that is, the CO/RC ratios
from all sources are constant to within a dispersion of a factor 
of 1.5.  These results 
emphasize both the small scatter in \qco\
within each galaxy as well as among different sources.
The CO and RC emission are on average linearly related for these
sources, though we
note that larger samples show a steeper RC/CO slope (${\S}$ 4.1).

Despite small-scale variations, there is no systematic trend of \qco\
with radius (lower-right panels of Figure 1).  In Figure \ref{figuraradial},
we show the correlations of all of the galaxies on a common plot.
The individual points represent annular averages and may generally
be read as radial profiles.  This figure serves to reinforce the
commonality of the CO-RC relation among the galaxies. We note that
this figure shows distance-independent intensity measurements, not luminosity 
measurements, and that the intensities extend over 3 orders of magnitude.

\begin{figure}[t]
\begin{center}
\includegraphics[width=10cm]{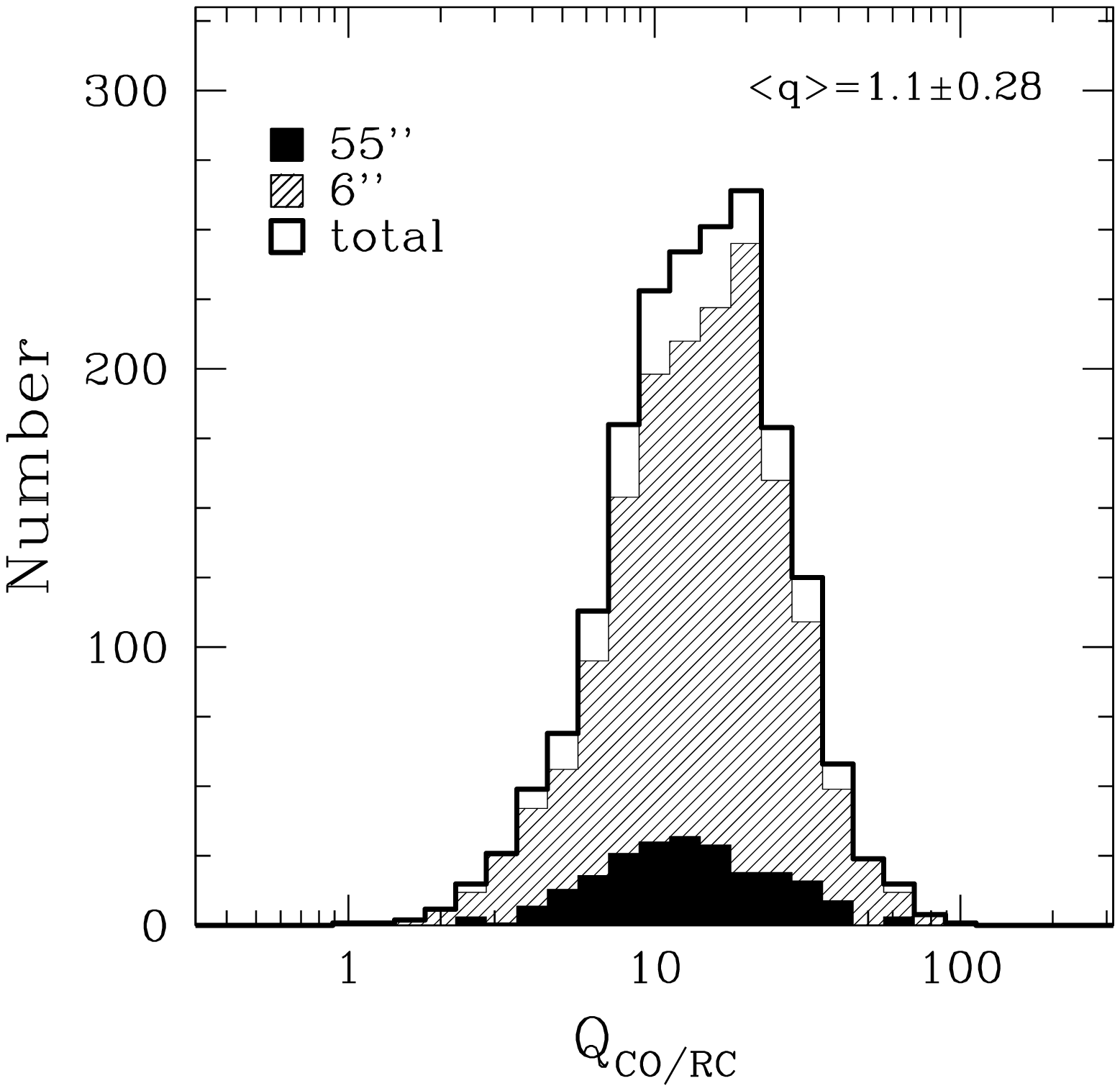}
\end{center}
\caption[] { Histogram of all measured \qco\ pixels.  The mean and dispersion
are also given. }
\label{figurahisto}
\end{figure}

Figure \ref{figuraflux} shows the correlation in the global \qco\
over entire galaxies.  Figure \ref{figuraradial} shows the correlation
at intermediate scales, when averaged over annular bins.
We can examine the scatter in \qco\ on the smallest scales by 
placing the individual pixels from all sources on a common plot, which we
present in Figure \ref{figurahisto}.  Over all pixels, the average
correlation is $<$\qco$>$ = 1.1 $\pm$ 0.28.

Some of the galaxies show slightly depressed
CO emission relative to the RC at their centers, which may be
caused by enhanced RC emission from the presence of an AGN. But with the exception of IC 342, the
deepest depression in the emission ratios is only a factor of 1.1--3.3 away
from the mean for each source.  This close agreement in
the galaxy centers is somewhat surprising, given the likely
variation in the CO-to-H$_2$ conversion in these regions. 

To summarize, we find a very good point-to-point correlation between the CO and RC
emissions at an angluar resolution of 6\arcsec~in a sample of 9 nearby spiral galaxies.
We are aware that the small size of the sample analyzed in this work could 
weaken the statistical relevance of this result. However, for a sample of 22 galaxies 
that includes an additional 13 sources, Paladino et al. (in preparation) find 
a similarly low scatter in the overall RC-CO correlation.

\subsection{ Variations in \qco\ With Spiral Structure}

To investigate trends as a function of location within the galaxies,
we also made images of the ratios \qco\ for each source 
(Figure \ref{figuraq}).  All images are shown on a common color scale,
so that color variations from source to source indicate real
variations in \qco.  

There are clearly some differences in the distribution of \qco\
among the different sources.  For NGC 4414, NGC 5005, and NGC 5033,
there is very little variation in \qco, as evidenced by
their low dispersions (Fig. 1).  It is worth noting
that both NGC 4414 and NGC 5005 are flocculent galaxies (Elmegreen
\& Elmegreen 1987), which lack prominent spiral structure.
In NGC 4736 and NGC 5194 (M~51),
systematic large-scale variations are apparent that follow 
the galaxies' spiral structure.  This is seen
in NGC 6946 as well, with enhanced \qco\ along the central
north-south elongated structure (variously interpreted as
a bar or as inner spiral arms; see Regan \& Vogel 1995) and in
the somewhat wispy arms to the southwest.  In NGC 5055,
there is a large-scale ringlike enhancement in \qco;
remarkably, the enhancement in this flocculent prototype
appears to follow 
nearly exactly two inner spiral arms revealed in $K\arcmin$ imaging
(Thornley 1996).  The most significant, large-scale
structures in \qco\ therefore appear to be organized along spiral
arms.  

\begin{figure*}[t]
\begin{center}
\includegraphics[height=9cm]{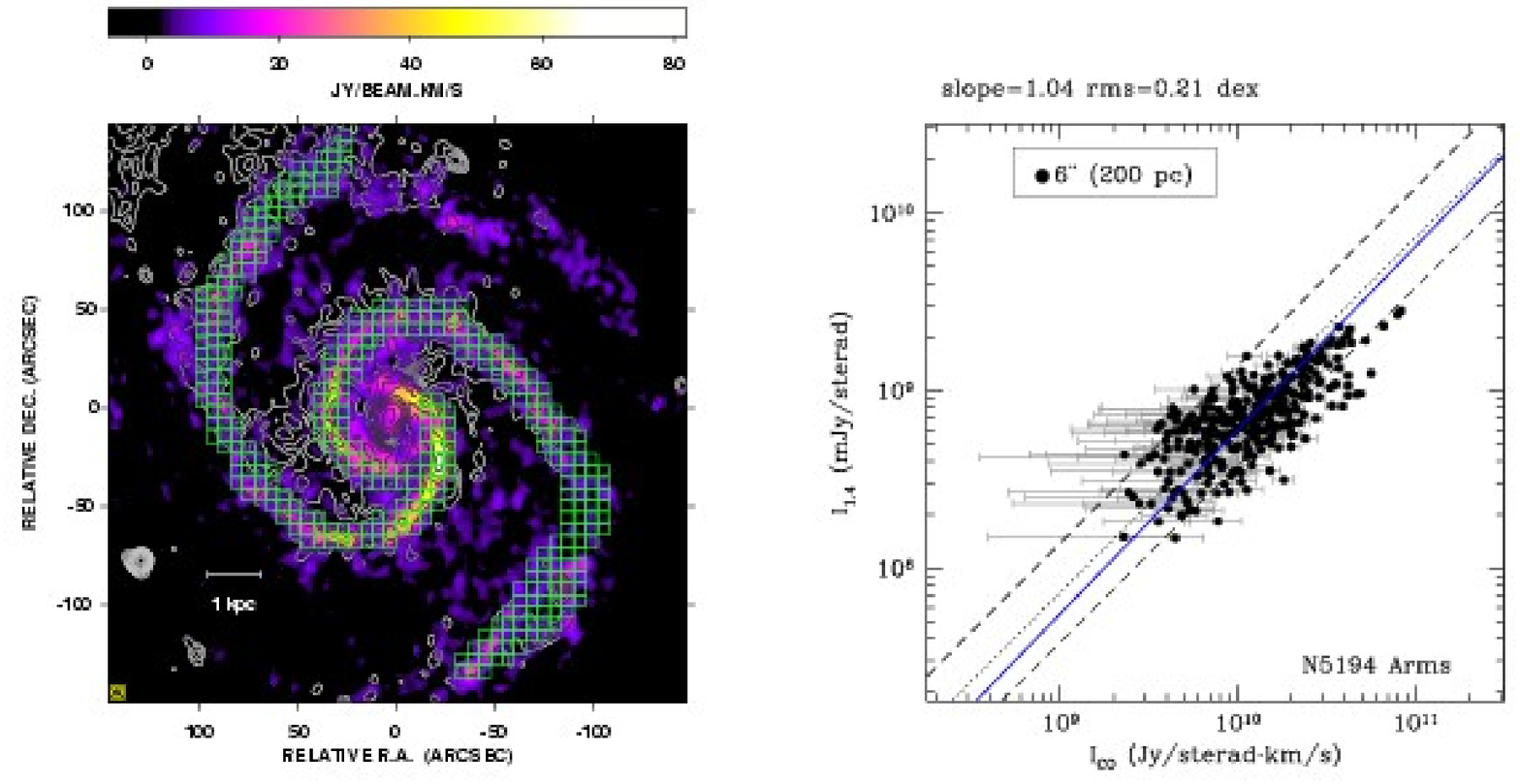}
\end{center}
\caption[] {Spatially resolved CO-RC correlation within NGC 5194 spiral arms. The RC and CO intensities are measured at 
independent positions in boxes of about 200$\times200$ pc in size. Spiral arms show a RC-CO correlation with a slope 
of 1.04 and with a scatter which is less than a factor of 2 . The RC/CO ratio in the arms is \qco$=1.2\pm0.20$, while in the interarm regions 
\qco$=1.0 \pm0.25$. }
\label{m51arms}
\end{figure*}

The two galaxies with the highest average \qco\ and among the
highest dispersion in \qco\ are IC 342 and NGC 4258.  The
variations in \qco\ in these galaxies do not appear to be closely
correlated with spiral structure; however, these galaxies are
unusual for other reasons, discussed further in Appendix A.

The association of higher CO/RC ratios on 1-2\,kpc scales in spiral 
arms was noted by Adler et al. (1991), who also measured a larger scatter
in the CO/RC relationship in these regions.
Given the improved resolution of our images, we performed a further 
analysis of the RC-CO correlation behavior in the spiral arms
of M51 with the intent of determining whether this systematic variation 
in \qco\ might correspond to a breakdown in the correlation.
We emphasize that we are sampling the RC and CO intensities at a
spatial resolution of about 200 pc, a value which is significantly below
the electron diffusion scale length. The result of this investigation 
is shown in Fig.~\ref{m51arms}. 
M51 still shows a tight linear RC-CO correlation, with a scatter of only 
0.2 dex, within the arms.
Based on this result, we cannot conclude that correlation breaks down at the 
linear scale of an arm width.  
If we consider the distribution of the CO/RC ratios in the arms, 
we get \qco\ $=1.2 \pm 0.2$,
which is only slightly above the average of $1.1 \pm 0.28$ that we get for
the entire galaxy sample. In the interarm regions, we measure
\qco\ =$1.0 \pm 0.25$.  On top of this systematic effect, which
appears to be associated with spiral structure, we observe larger 
but isolated deviations in \qco\ which contribute to the correlation scatter.
However, despite these localized variations there is still a remarkably
low dispersion in the high-resolution CO-RC correlation.
To summarize, we tend to measure an increase in CO/RC in spiral arms
as seen earlier in Adler et al. (1991); unlike Adler et al.,  we do not see 
an increase in scatter within the spiral arms: the correlation dispersion 
is just as low.

\begin{table*}
\caption{Global Fluxes and Scale Lengths in BIMA SONG Sources}
\begin{center}
\begin{tabular}{llllllll}
\noalign{\smallskip}
\hline
\noalign{\smallskip}
Galaxy & S$_{1.4}$ & S$_{60}$ &  S$_{CO}$ & Reference$^{~a)}$  & $l_{CO}^{~~b)}$ & $l_{RC}^{~~c)}$ &  $l_{CO}$/$l_{RC}$ \\
     & (mJy)    &  (Jy)  &  (Jy~km~s$^{-1}$) & &  (kpc)    &  (kpc)   & \\
\hline
IC 342   & 2475 & 255.96 & 20000 & 1,2,3  & 3.5 $\pm$ 0.2 & 6.21 $\pm$ 0.06 & 0.56\\
NGC 0628 & 180  & 20.86  & 1514  & 2      & 3.2 $\pm$ 0.1 & 15.3 $\pm$ 0.1 & 0.21\\
NGC 1068 & 4991 & 181.95 & 4102  &        & 1.6 $\pm$ 0.4 & 0.7 $\pm$ 0.2 & 2.16\\
NGC 2903 & 407  & 60.03  & 3254  &        & 2.0 $\pm$ 0.2 & 1.7 $\pm$ 0.2 & 1.21\\
NGC 3351 & 48   & 19.92  & 1513  &        & - & -&- \\
NGC 3521 & 357  & 47.02  & 4800  &        & 2.1 $\pm$ 0.2 & 4.1 $\pm$ 0.1 & 0.52\\
NGC 3627 & 458  & 67.8   & 4259  &        & 5.1 $\pm$ 0.1 & 3.5 $\pm$ 0.1 & 1.45 \\
NGC 3938 & 62   & 9.24   & 923   &        & 4.0 $\pm$ 0.2 & 6.3 $\pm$ 0.1 & 0.64 \\
NGC 4258 & 792  & 21.60  & 2686  & 1,2    & 2.1 $\pm$ 0.2 & 3.1 $\pm$ 0.3 & 0.70 \\
NGC 4303 & 416  & 37.53  & 2427  &        & 3.2 $\pm$ 0.1 & 3.3 $\pm$ 0.1 & 0.96 \\
NGC 4321 & 340  & 25.86  & 2972  &        & 4.8 $\pm$ 0.1 & 5.0 $\pm$ 0.1 & 0.96 \\
NGC 4414 & 231  & 30.11  & 2453  &        & 3.4 $\pm$ 0.2 & 3.00 $\pm$ 0.03 & 1.13 \\
NGC 4569 & 83   & 10.08  & 1096  &        & 2.6 $\pm$ 0.1 & 2.5 $\pm$ 0.2 & 1.04 \\
NGC 4736 & 254  & 69.2   & 2641  &        & 0.9 $\pm$ 0.4 & 0.7 $\pm$ 0.2 & 1.22 \\
NGC 4826 & 103  & 35.45  & 1845  &        & 0.6 $\pm$ 1.1 & 0.4 $\pm$ 0.7 & 1.46 \\
NGC 5005 & 176  & 22.3   & 1278  &        & 2.9 $\pm$ 0.3 & 3.2 $\pm$ 0.1 & 0.90 \\
NGC 5033 & 178  & 16.45  & 2469  &        & 4.5 $\pm$ 0.1 & 3.2 $\pm$ 0.1 & 1.39 \\
NGC 5055 & 390  & 40.02  & 3812  &        & 2.4 $\pm$ 0.1 & 2.3 $\pm$ 0.1 & 1.03 \\
NGC 5194 & 1490 & 98.8   & 10097 &        & 2.9 $\pm$ 0.1 & 3.4 $\pm$ 0.1 & 0.84 \\
NGC 5247 & 53   & 8.07   & 1102  & 4,5    & 7.1 $\pm$ 0.2 & 7.3 $\pm$ 0.1 & 0.97 \\
NGC 5248 & 140  & 20.71  & 1829  &        & 3.0 $\pm$ 0.3 & 4.2 $\pm$ 0.1 & 0.72\\
NGC 5457 & 750  & 88.04  & 3479  &        &  - & -& -\\
NGC 6946 & 1395 & 136.69 & 9273  & 2,4    & 2.5 $\pm$ 0.2 & 3.83 $\pm$ 0.04 & 0.65\\
NGC 7331 & 373  & 35.29  & 2762  & 2,4    & 3.2 $\pm$ 0.1 & 3.33 $\pm$ 0.03 & 0.96\\
Average  &      &        &       &        &               &                 & 0.98 $\pm$ 0.41\\
\hline
\noalign{\smallskip}
\label{globalfluxes}
\end{tabular}
\end{center}
\begin{list}{}{}
\item[$a)$] Flux references: (1) S$_{1.4}$ is 1.4 GHz RC from White \& Becker 1992; 
(2) S$_{60}$ is 60 $\mu$m emission from Rice et al. 1988; 
(3) S$_{CO}$ from Crosthwaite et al. 2001; 
(4) S$_{1.4}$ from Condon 1987.  
(5) S$_{60}$,S$_{100}$ from IRAS Faint Source Catalog;
If not otherwise indicated, S$_{1.4}$ is from Condon et al. 1990, 
S$_{60}$ is
from the IRAS Bright Galaxy Sample, Soifer et al.  1989, and
S$_{CO}$ is from Helfer et al. 2003.
\item[$b)$] CO scale lengths modeled from the 55\arcsec\ 12 m data.
\item[$c)$] RC scale lengths modeled from the low-resolution RC data from either the present 
study or from NVSS imaging.
\end{list}
\end{table*}

\subsection{The CO, RC and FIR Length Scales}

We have modeled the CO and RC exponential scale lengths from
22 BIMA SONG galaxies with CO data taken from the 55\arcsec\
12 m data and the low-resolution RC data from either the present 
study or from NVSS imaging.  The results are listed in 
Table~\ref{globalfluxes} and shown in Figure~\ref{lengthshisto}.  
We note that the formal errors in 
the scale lengths almost certainly underestimate the true errors, 
for the following reasons:  (1) the images often cover only
1--2 $e$-folding lengths  in the galaxies; (2) given
the small coverage, the radial profiles may or may not be fairly 
represented by exponential functions.  At higher resolution, the
CO radial profiles have much more detailed structure and are in
general not well characterized by exponentials (Regan et al. 2001);
(3) there may be systematic errors in the CO data that are not
well represented by the formal errors (Helfer et al. 2003);
(4) we have not excluded background radio continuum sources; this
error might tend to produce longer RC scale lengths than 
are present intrinsically.
We note that some of the CO scale lengths derived here differ
from those published in Regan et al. (2001).  
The scale lengths
presented here are to be preferred (Regan 2004, private communication).

Despite the qualifications (which may also apply to
previously determined FIR and RC scale lengths), the data in 
Table~\ref{globalfluxes} suggest
that on average, the CO scale lengths are comparable to the
RC scale lengths, with $<l_{CO}/l_{RC}>$ = 0.98 $\pm$ 0.41.  
This is true even though the dispersion in 
the average is large; that is, some galaxies show significantly 
longer or shorter CO disks than RC disks.   
We can further compare the FIR scale lengths modeled by
Marsh \& Helou (1995) with the BIMA SONG CO disk scale 
lengths for an overlap sample of 5 galaxies: NGC 2903, NGC 4303,
NGC 5055, NGC 5194, and NGC 6946.   (We exclude NGC 628,
which lacks RC in the central region where the CO is present;
the fit to the ringlike RC distribution yields a large scale length
which is poorly determined.  We also exclude NGC 7331,
with ringlike emission in CO and RC that also precludes a good
determination of its scale length.)
After scaling the Marsh \& Helou (1995) $l_{60\mu m}$ to a
common distance scale with the BIMA SONG study, we find that
for this subsample,
$l_{60\mu m}/l_{CO}$ = 0.69 $\pm$ 0.07,
$l_{60\mu m}/l_{RC}$ = 0.65 $\pm$ 0.14, and
$l_{CO}/l_{RC}$ = 0.94 $\pm$ 0.20.
These results suggest that the FIR tends to have a smaller scale length
than the RC or CO and that the RC and the CO may in some sense
be more directly related to each other than either is to the FIR.
New determinations of the FIR scale lengths using {\it Spitzer} imaging
should help to clarify this issue.

\section{Discussion}

\subsection{ The Radial Correlations}

The most significant deviation from the close correlation between
the FIR and RC emission is a small, generally monotonic decrease
in FIR/RC as a function of radius in many extragalactic sources
(Beck \& Golla 1988; Rice et al. 1990; Bicay \& Helou 1990).
In a seminal study, Marsh \& Helou (1995) studied fully 2-dimensional
maps of the FIR and RC emissions from 25 nearby
galaxies, with improved, $\sim$ 1\arcmin\ resolution over previous
studies.  Their results show decreasing
ratios of FIR/RC in nearly all sources, with disk values
at $\sim$4\arcmin\ radius (10--20 kpc) 
that are lower by up to a factor of
10 than the central ratios.  In the Marsh \& Helou study,
there are 3 galaxies in common with the present 
study: NGC 5055, NGC 5194, and NGC 6946.  These 
galaxies show monotonic decreases in FIR/RC of a factor of 3.7, 4.1, 
and 2.8 over 
4--5\arcmin\ in radius.  By contrast, the galaxies in
the present study, including these three galaxies, do
not show any apparent trend of CO/RC with radius (see bottom
right panels of Figure 1).
The CO-RC correlation therefore seems to be different from the
FIR-RC correlation:  the CO/RC shows no systematic trend with radius, 
and the CO and RC scalelengths seem to be on average equal, with both
longer than the FIR scalelengths (${\S}$ 4.4).

In the CO-RC correlation, we have seen that the strongest organized
structures in \qco\ are found along spiral arms, with enhanced
\qco\ along the spiral arms and lower values in the interarm regions.
Recent {\it Spitzer} FIR imaging of M~81 (Gordon et al. 2004)
shows very similar features:  first, that the
variations in \qfir\ are organized on size scales
far greater than the resolution of the observations; and
second, that these large-scale variations in \qfir\ strongly follow
the spiral structure in M~81.   In fact, though
Gordon et al. don't discuss this issue, an examination of the \qfir\
map of M~81 appears to show no systematic radial trend in \qfir.  

How can the M~81 FIR-RC result and the CO-RC results presented in
this study,
namely that the strongest variations in the ratios are seen in
azimuth across
spiral arms, not as a radial variation, be justified with 
IRAS- and ISO-era observations that showed decreasing FIR-RC as a function 
of radius?   If the strongest intrinsic variations are indeed
across spiral arms, we note that the effect of these variations will
generally be largest at small radius.  That is, when averaged in
azimuthal rings, the fraction of the total area in an annulus
taken up by the spiral arms is much larger at small radius than it
is at large radius.  It may be that the
measured radial falloff in FIR-RC is primarily a consequence of the
low resolution of the FIR observations; it is essentially a
measurement of the arm-interarm contrast of the galaxy.
We predict that high-resolution {\it Spitzer} observations from the
SINGS team (Kennicutt et al. 2003) and others will 
show that M~81 is not anomalous in this sense:  the dominant variation 
will also be across spiral arms, with any radial trend being a
secondary effect.  
In any case, {\it Spitzer} imaging will certainly further 
illuminate the issue of the relative strengths of the CO-RC,
FIR-RC and the CO-FIR correlations.

\subsection{A Brief Summary of Models of the FIR-RC and CO-RC Correlations}

The global FIR-RC correlation is valid over five orders of magnitude in
luminosity (Yun et al. 2001).  We have shown in this paper that the 
RC, FIR and CO emission are also globally correlated,
and that the CO-RC correlation is equally good on scales down 
to several hundred parsecs.  
However, the physical bases for understanding the FIR-RC and CO-RC
correlations are not well understood. 
The problem is that the emission mechanisms 
are so different that it is difficult to understand why such a tight 
relationship should hold between them.  The FIR emission 
is thought to be from dust heated by UV radiation from hot young stars 
and is therefore a reasonably good tracer of the 
star formation rate in a galaxy 
(e.g. Devereux \& Young 1991).  Because the extinction in the UV is so high,  
essentially all of the UV photons are absorbed by dust and are reprocessed 
into FIR emission. Few Lyman continuum photons are thought to escape from a 
galaxy (Leitherer et al. 1995; Heckman et al. 2001; Deharveng et al. 2001).  
Over scales of a few hundred parsecs or more, the RC 
emission from galaxies is typically at least 90\% non-thermal (Condon
1992); for the galaxies in this paper, the thermal fractions at 1 GHz
are typically 5--10\% (Niklas, Klein \& Wielebinski 1997).  This 
synchrotron emission is due to cosmic-ray electrons 
spiraling around the galactic magnetic field. Why should the emission from 
relativistic cosmic-ray electrons be so tightly correlated with the FIR 
from dust?  The CO emission from giant, turbulent, cloud complexes 
traces the bulk of molecular gas 
mass, even though it is optically thick. 
The good correlation between the CO and the 
FIR has been known for some time and is generally attributed to the intimate 
connection between massive star formation and molecular clouds (Devereux 
\& Young 1991).  
That is, because CO has been found to be a reliable 
tracer of the mass of GMCs, and because the rate of massive star formation 
is related to the mass of molecular gas through an approximately constant 
star formation efficiency averaged over time (Rownd \& Young 1999), the 
FIR, through the UV production and star formation rates, should be 
coupled to the CO emission.  In detail, however, there
is scatter of at least an order of magnitude in the
CO-FIR relation (Young \& Scoville 1991), and it has
been difficult to relate the FIR reliably to a star formation
rate for different types of galaxies (e.g. Kennicutt 1998; Bell 2003).

In the following, we briefly review some of the models 
previously posed to explain the FIR-RC and CO-RC correlations.

\subsubsection{The FIR-RC Correlation: Conventional Picture and Calorimetric 
Models}

In the conventional model, the FIR and the RC are 
correlated because each is a tracer of massive star formation: the RC through 
the supernovae that are ultimately produced from the recently formed 
massive stars and the FIR from the dust-reprocessed UV radiation from 
the stars themselves.  But this explanation is wanting on several accounts.  
First, why should the density of cosmic-ray electrons and the strength of 
the interstellar magnetic field be closely coupled to the rate of star 
formation?  
The production of the RC emission requires a series of steps: massive 
star formation $\to$ supernova production $\to$  
cosmic-ray acceleration $\to$ 
interaction with the magnetic field, as does the FIR emission: massive 
star formation $\to$ UV photon production $\to$
absorption and reradiation 
by dust.  To get the FIR-RC correlation to hold to a factor of two, as 
observed, each step must have a smaller range of variation
than the observed correlation, 
which seems implausible.  Furthermore, since the FIR-RC correlation 
is observed over a large range of physical scales, it is unclear why it 
should be so good on galactic scales, since star formation is a local 
process.  In addition, star formation is seen to correlate well with 
{\it thermal} RC emission in HII regions (Boulanger \& Perault 1988;
Haslam \& Osbourne 1987; Wells 1997), 
rather than the non-thermal 
emission on scales less than 100 pc, and there is not a particularly 
good correlation of supernova remnants with either sites of star 
formation (Wootten 1978) or with mid-IR or FIR emission (Cohen \&
Green 2001; Whiteoak \& Green 1996).

Another problem with the conventional model is that the CO and RC 
both have a somewhat poorer correlation with other SFR indicators
than they do with each other; indeed Bell (2003) raises 
issues about the linearity of the FIR-RC correlation by arguing that 
both indicators underestimate the star formation rates at low luminosities.
Bell goes on to reach the unlikely conclusion that two unrelated effects
have conspired to generate the observed correlation, without questioning 
the fundamental assumption that it is the star formation rate that 
provides the key link to the FIR-RC correlation. 
Leroy et al. (2004) show that the CO-RC correlation
holds even for low-luminosity dwarf galaxies, suggesting that even further
ad-hoc tuning would be necessary in order to understand the correlations 
among the CO and the FIR and RC within a conventional SFR framework.
Notwithstanding Bell's arguments regarding the
suitability of the FIR and RC as star formation tracers, 
the correlations among the FIR-RC and CO-RC remain tight
{\it even} if the FIR and RC underestimate the star formation, and we
must look to an alternate explanation to understand the
observed correlations.

V{\"o}lk (1989) and others since have presented a calorimetric
model in which the ratio of RC and FIR emissions
is held constant by assuming that both the relativistic electrons
as well as the UV photons are proportional to the SN rate.
In this scenario, all systems are optically thick, so that
all synchrotron or inverse Compton losses must occur in the host
galaxy, and all the FUV radiation from massive stars is absorbed
by dust grains within the galaxy.  
With both of these processes driven by the massive stars, the
galaxy acts as a calorimeter.  

Niklas \& Beck (1997)
show that only $\sim$ 30\% of their 74-galaxy sample have
radio spectral indices that support the optically thick model;
however, Lisenfeld \& V{\"o}lk (2000) argue that cosmic ray
electron convection, rather than diffusion, can flatten the
spectra within the calorimeter model.
Finally, the calorimeter model addresses the global energetics within
galaxies, and it does not address the observed FIR-RC correlation that holds 
at least on kiloparsec size scales within galaxies, or the CO-RC 
correlation that holds down to size scales of $\sim$ 100 pc.  This
model thus seems to be unsatisfactory as a solution that holds
generally for different kinds of galaxies at various resolutions.

In the phenomenological model by Helou \& Bicay (1993), 
the FIR-RC correlation is driven by a tight coupling
between the dust-heating photon luminosity and the
production rate of cosmic ray electrons.  The photon
luminosity is regulated by the effective optical depth
to the photons, and the CR luminosity is regulated by 
the efficiency with which the cosmic rays emit synchrotron
radiation before they escape.  In the optically thick case, the model
is similar to calorimetric models, which Helou \& Bicay postulate 
applies to infrared-luminous galaxies.  In the optically thin case,
which Helou \& Bicay apply to normal galaxies, the photon luminosity
and the CR luminosity  are predicted to be approximately
equal because of a coupling of the magnetic field with the
gas density.  Niklas \& Beck (1997) however point out that in this
model, the transition between galaxies with low and high radiation fields
should be accompanied by an increase in the nonthermal
spectral index, which they do not observe.
Furthermore, Niklas \& Beck show that their
measured spectral indices do not correlate well with the
equipartition magnetic field strength,
in contradiction with the Helou \& Bicay (1993) picture.

\subsubsection{The CO-RC correlation: Cosmic Ray Heating and Secondary 
Electrons}

Suchkov, Allen \& Heckman (1993) present a model for the
CO-RC correlation where the nuclear component of cosmic rays
is responsible for heating the molecular
gas while the electron component account for the synchrotron radiation.
In this model, the CO need not be directly tracing star
formation; rather, its emission depends on the cosmic rays that are
products of high-mass star formation.
This model is particularly attractive since it naturally provides a
direct physical link between the CO line integrated intensity and the
RC synchrotron emission.
Bradford et al. (2003) show that the bulk of the molecular gas
in the central 180 pc of NGC 253 is highly excited; these authors infer
a temperature of $T\sim 120$ K for the molecular gas and conclude that
the best mechanism for heating gas is cosmic rays that are produced
towards the starburst nucleus of this galaxy.
However, we show that the observed CO-RC correlation spans more than
three orders of magnitude in both quantities. This poses a serious problem
for the CR heating model in that the molecular gas temperature should be raised
by at least a factor of 1000. Thus, if it cannot be excluded that
the CR heating is acting in some circumstances, it is difficult to
accept it as the general explanation for the CO-RC correlation.

An interesting aspect of this picture, although not often
addressed in literature, is the issue of the secondary electrons, which
are produced by CR primaries interacting with the molecular gas
(e.g. Marscher \& Brown 1978). If the production of the secondary electrons
is significant, then the CR nuclei could not only raise the molecular gas
temperature but also enhance the synchrotron radiation. As outlined
in ${\S}$ 1, the RC emission from isolated giant molecular clouds is mostly
thermal. Thus, the secondary electrons able to survive collisional losses will probably 
escape molecular clouds in a very short time-scale and
 diffuse in the disk where they became indistinguishable from the primaries. 
One can devise a scenario in which the CO-RC correlation is governed by the volume density 
of molecular clouds.
The CO integrated intensity traces the number of clouds within the 
volume of the galaxy disk intercepted by the beam.  On other hand, the 
secondary electron injection will also be proportional to the volume 
density of clouds, leading to a linear CO-RC correlation.
A precise calculation of the secondary electron contribution
implies special choices of rather uncertain parameters, such as the
efficiency of energy transfer from supernovae to CRs and the ratio 
of primary electrons to nuclei in CRs (see e.g. Paglione et al. 1996) or
the effective amount of diffuse molecular gas in the interstellar medium.
While a detailed discussion of this issue is beyond the scope of
this paper, such an intertwined relationship may help to explain 
the closeness of the CO and RC emission.

\begin{figure*}
\begin{center}
\includegraphics[width=17cm]{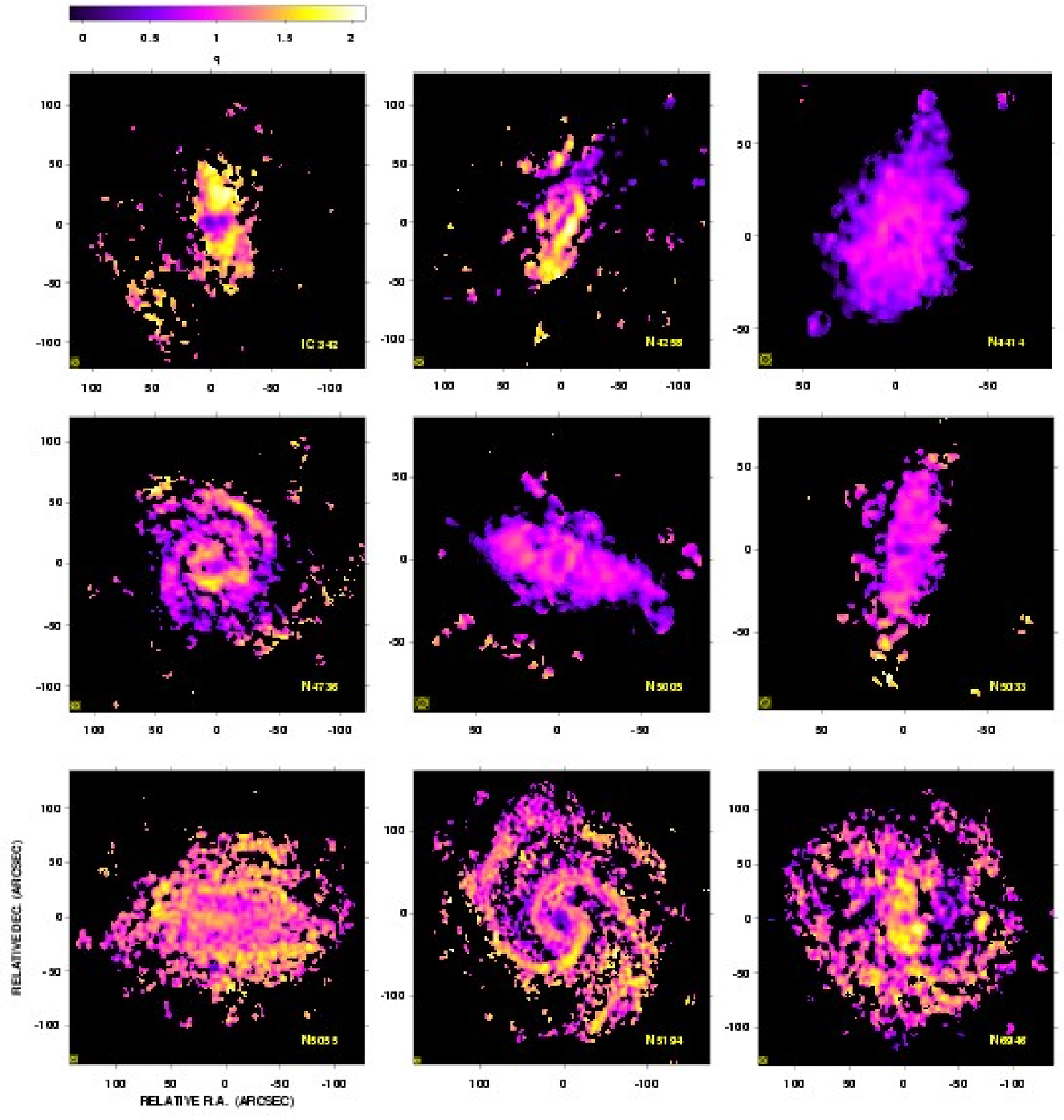}
\end{center}

\caption[] {Images of \qco\ as defined in ${\S}$3. The scale is given
as a color wedge at the top of the figure.  }
\label{figuraq}
\end{figure*}

\subsubsection{Cosmic-Ray Driven Interstellar Chemistry}

Bettens et al. (1993) proposed that chemical evolution in
molecular clouds, driven by the cosmic ray flux, is linked
to the FIR flux through ion destruction in the clouds.  
While taking into consideration all three of the RC, FIR, and 
CO emissions, this theory neglects the strong influence of the 
interstellar magnetic field on the synchrotron radiation.  The theory 
also makes rather specific predictions about how the relative
abundances of molecules like HCN, H$_2$CO, and CH$_3$OH are
tied to the nonthermal RC.  Based on results from this paper and
from how the HCN/CO ratios vary within the Milky Way and other
galaxies (e.g. Helfer \& Blitz 1997$a$;$b$), it seems
unlikely that these predictions will hold in detail.

\subsection{Hydrostatic Pressure as a Regulating Mechanism for the 
CO-RC Correlation}

The difficulties in relating the seemingly independent emission
mechanisms, and doing so over the range of size scales observed,
suggest that there might be a single physical parameter that acts to
correlate the observed quantities, rather than some process that
relates them directly.  We suggest that the parameter is the midplane
hydrostatic pressure $\bar{P}$, modified by other large scale
pressure variations such as those produced by the spiral arms.
Recently, Wong (2000), Wong \& Blitz (2002) and Blitz \& Rosolowsky
(2004) have suggested that the azimuthally averaged fraction of gas in
the molecular phase, $f_{mol}$, or equivalently, the H$_2$ to HI
ratio, is determined by the radial variation in interstellar gas
pressure, which can be approximated as

\begin{equation}
\bar{P} = 0.84 (G \Sigma_*)^{0.5}\Sigma_g \frac {v_g} {(h_*)^{0.5}} \;.
\label{approxpressure}
\end{equation}

\noindent Here $\Sigma_{g}$ is the total surface density of gas,
$\Sigma_{*}$ is the surface density of stars, $v_g$ is the velocity
dispersion of the gas, and $h_*$ is the stellar scale height, which
can be written in terms of the stellar velocity dispersion $v_*$ and
midplane volume density $\rho_*$ as $h_* = ({v_*}^2/4\pi G
\rho_*)^{0.5}$.

This expression for $\bar{P}$ assumes that the midplane volume density
of stars greatly exceeds the midplane volume density of gas, a
condition that is satisfied throughout most disk galaxies except
perhaps in spiral arms.  One particular utility of this formulation is
that both $\Sigma_{g}$ and $\Sigma_{*}$ are directly measurable, while
$v_g$ and $h_*$ are observed to be fairly constant within galaxies
($v_g$: Shostak \& van der Kruit 1984; Dickey et al. 1990; $h_*$: van
der Kruit \& Searle 1981a; 1981b; Fry et al.  1999).  Furthermore,
$v_g$ shows little variation from galaxy to galaxy about a value of
$\sim$8 km s$^{-1}$, and $h_*$ has a dispersion of only 50\% from the
mean (Kregel, van der Kruit \& de Grijs 2002).  In any event the
dependence of the pressure on $h_*$ is weak due to the exponent of 0.5.

Let us first consider the origin of the CO-RC correlation.  
We begin by making the usual assumption that the CO surface
brightness scales linearly with $\Sigma_{\rm H_2}$.
Observations show that $\Sigma_{\rm H_2}/\Sigma_{\rm HI} \propto 
\bar{P}^{0.8-0.9}$ (Wong \& Blitz 2002; Rosolowsky \& Blitz in preparation).
Because the surface density of atomic hydrogen $\Sigma_{\rm HI}$ is
roughly constant with radius (e.g. Wong \& Blitz 2002), we find that

\begin{equation}
I_{\rm CO} \propto \bar{P}^{0.8}\;.
\label{copress}
\end{equation}

\noindent The shallow decrease of $\Sigma_{\rm HI}$ with radius in
some galaxies would increase the exponent slightly.

The RC emission, on the other hand, is determined largely by the
synchrotron emissivity, which can be written as

\begin{equation}
I_{\rm RC} \propto N_0 B^{(\gamma +1)/2} \nu^{(1-\gamma)/2}\;,
\label{syncflux}
\end{equation}

\noindent where $B$ is the magnetic field strength, $N_0$ is the
number density of cosmic-ray electrons with energies between energy
$\epsilon$ and $\epsilon + d\epsilon$, and $\gamma$ is the power law index of the spectrum
of cosmic-ray energies such that $N(\epsilon) = N_0 \epsilon^{-\gamma}$ (Condon
1992). The power law index of the observed synchrotron spectrum
suggests that typically $\gamma = 2.6$ at GeV energies; thus we
expect $I_{\rm RC} \propto N_0 B^{1.8}$.

To express $B$ in terms of $\bar{P}$, we assume energy equipartition
between the turbulent energy which dominates the hydrostatic pressure
$\bar{P}$ and the magnetic energy density $B^2/8\pi$, so that $B
\propto \bar{P}^{0.5}$.  This assumption has been discussed previously
in the literature with regard to the FIR-RC correlation (e.g. Helou
\& Bicay 1993; Groves et al.\ 2003).    We therefore expect
$I_{\rm RC} \propto N_0 \bar{P}^{0.9}$.

To express $N_0$ in terms of $\bar{P}$, we consider the radial distributions   
of both quantities as observed in the Milky Way.         
The radial distribution of $N_0$, although uncertain,
is based on gamma-ray data from the Compton Gamma Ray Observatory,
and yields a
power-law fit: $N_0 \propto R^{-0.56}$ for R $\ga 3$ kpc (Strong \&
Mattox 1996). The radial distribution of $\bar{P}$ is shown in Figure
\ref{mwpressure}, using the data of Dame (1993) for $\Sigma_g$ and an
exponential distribution of stars with a radial scale length of 3 kpc
(Spergel, Malhotra \& Blitz 1996) tied to a value of $\Sigma_*$ = 35
M$_{\sun}$ pc$^{-2}$ at the solar circle.  Beyond 4 kpc, $\bar{P}$ is
seen to have an approximately power law form with $\bar{P} \propto
R^{-2.2}$.  Combining these two relations suggests that $N_0 \propto
\bar{P}^{0.25}$.  We therefore expect 

\begin{equation}
I_{\rm RC} \propto \bar{P}^{1.15}.  
\label{rcpress}
\end{equation}

\noindent
Combining Equations 2 and 4 yields

\begin{equation}
I_{\rm RC} \propto I_{\rm CO}^{~1.4},
\label{rcco}
\end{equation}

\noindent
which is consistent within the uncertainties with the global results
given in ${\S}$ 4.1 and with the intermediate-resolution results
of Murgia et al. (2002) for spiral galaxies and of Leroy et
al. (2004) for dwarf galaxies.

As a check on the above derivation of $N_{0}$, we note that
if we combine the above result $I_{\rm RC} \propto N_{0} \bar{P}^{0.9}$ with
Equation (2), we derive $I_{\rm RC} \propto N_{0} I_{\rm CO}^{~1.1}$.
For the large-scale observations presented here, in Murgia et al. 
(2002) and in Leroy et al. (2004), we have
$I_{\rm RC} \propto I_{\rm CO}^{~1.3}$, which suggests 
a scaling of the CR electron density with pressure
$\langle N_0 \rangle \propto  \langle \bar P \rangle ^{~0.16}$
(quantities in brackets represent values averaged over the entire galaxy).
For the sub-sample of the 9 bright galaxies studied at high resolution in 
this work, we instead observe
$I_{\rm RC} \propto I_{\rm CO}^{~1.1}$.
The sub-sample observations suggest that $N_{0}$ may be nearly independent 
from $\bar P$ for these galaxies; in any case, the intrinsic variation
from source to source suggests that $N_{0}$ could vary somewhat among 
galaxies.  However, given the small scatter of the
CO-RC correlation, it appears that $N_{0}$ varies only
slowly with $\bar P$.

\begin{figure}
\begin{center}
\includegraphics[width=8cm]{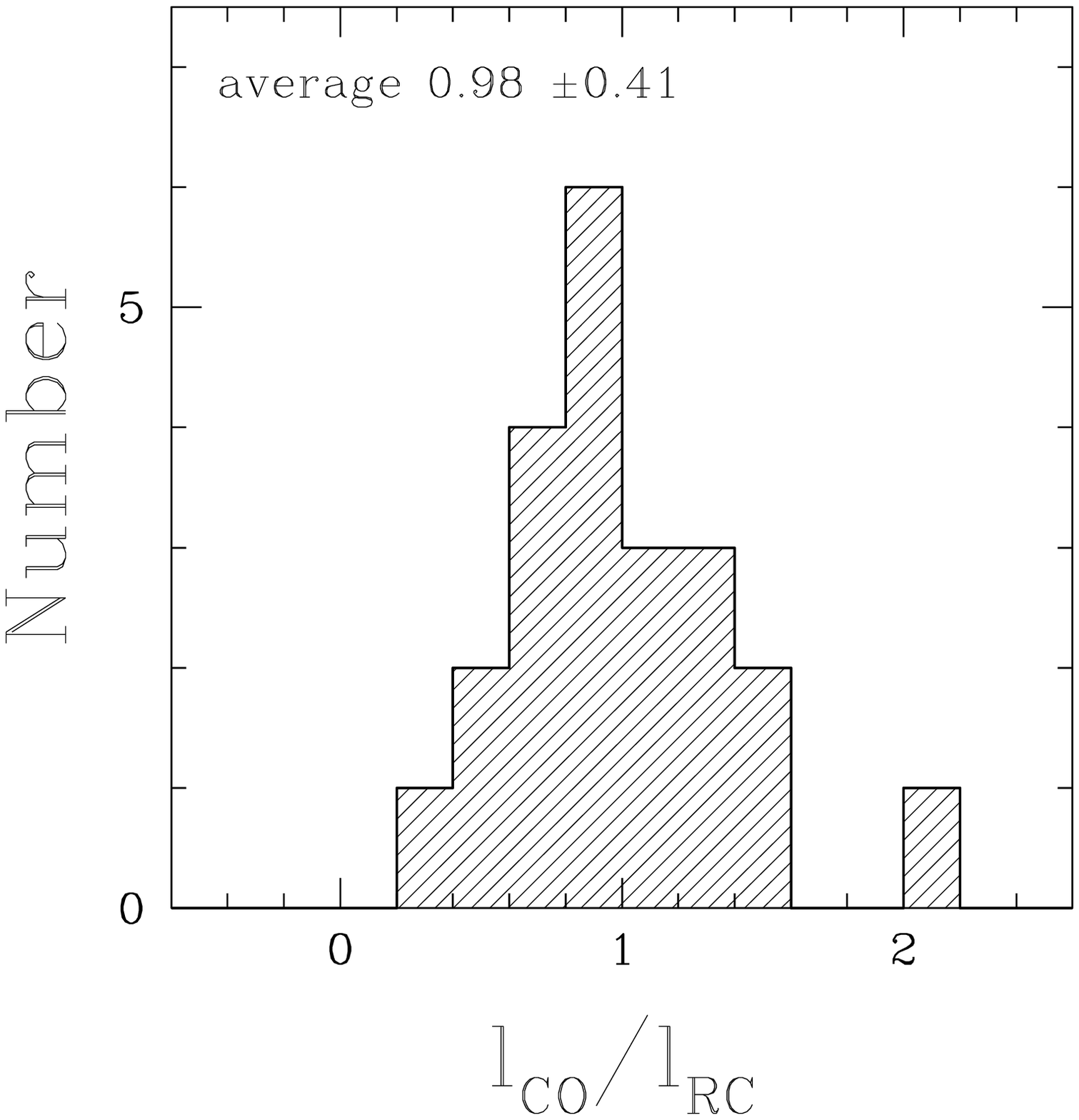}
\end{center}
\caption{Histogram of the CO to RC scale lengths ratio. The average 
 ratio considering all galaxies results $0.98\pm0.41$. 
}
\label{lengthshisto}
\end{figure}

Next we consider how the FIR emission is related to both the CO and RC.
The FIR emission has traditionally been linked to the CO emission
through a Schmidt-type star formation law, with the assumption that
the FIR is directly proportional to the star formation rate per unit area.
This formulation works reasonably well for luminous FIR galaxies
(Kennicutt 1998), but is problematic for normal and
low-luminosity sources (Kennicutt 1998; Bell 2003).  

Recently Dopita et al.\ (2004) proposed that the FIR emission 
itself is also directly related to $\bar{P}$:  the ISM pressure
sets the stall radius of the H\thinspace{\sc ii} region, so
that with higher $\bar{P}$, the H\thinspace{\sc ii} region
stalls at smaller radius and the mean dust temperature in the
atomic and molecular shell around the H\thinspace{\sc ii} region
should be higher. From stellar spectral synthesis modeling, Dopita et al. show that
the 60 $\mu$m flux is positively correlated with the ISM pressure,
though in their ``toy'' model the increase is not enough to account for the
corresponding increase in the RC emission.  However, taken together
with the above model of the CO-RC correlation, these models may
have the potential to link all three of the CO-RC-FIR emissions
with the interstellar pressure, without any {\it explicit}
dependence on a Schmidt-type law or other star formation scenario.
This is a particularly attractive feature given the problems with
fitting the details of a generic starforming model to high-
and low-SFR galaxies.

Why is it that the CO-RC-FIR relationship holds both globally and within
galaxies to scales of a few hundred parsecs, but breaks
down on smaller scales, scales of individual GMCs?  Pressure
mediation provides a natural scale for the relationship: the pressure
scale height in galactic disks.  On larger scales, hydrostatic
pressure is determined by the slow variation of the gravity provided
by the stars, on size scales equal to or larger than the thickness
of the gas layer in a galactic disk.  In the vicinity of
the Sun, this scale is about 250 pc, but the thickness
varies with galactic radius.
Superposed on this is a higher spatial frequency
variation in the spiral arms; on these and smaller scales, local
pressure effects can become important.

Thus, our theory suggests that it is primarily HS pressure that
drives the correlation by regulating both the fraction of gas in molecular
form and the magnetic field strength. Secondary effects that can
modify the basic correlation include (but are not limited to) spiral
density waves, local star formation effects, and CR propogation.

\begin{figure}
\begin{center}
\includegraphics[width=9cm]{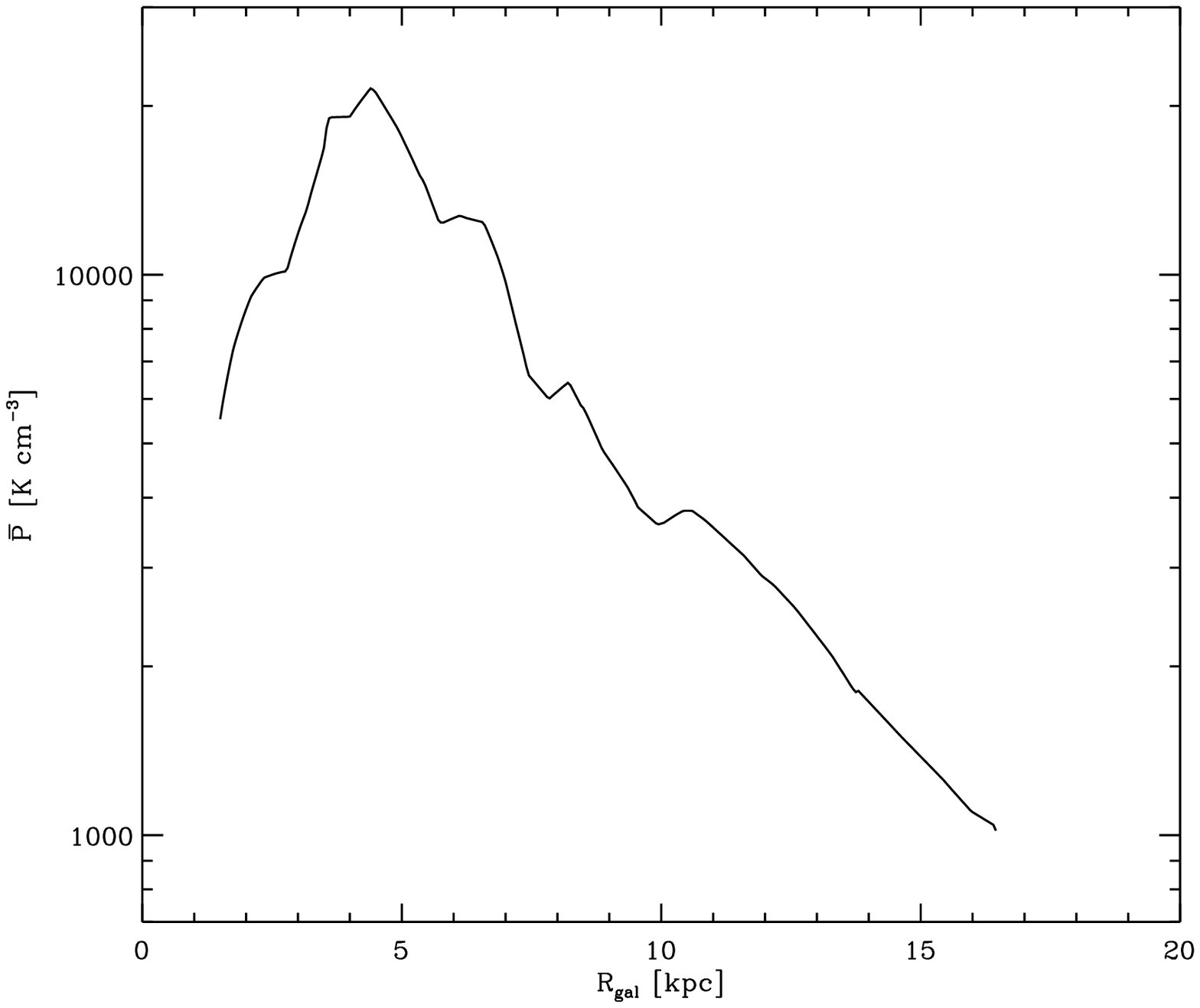}
\caption{$\bar P$ for the Milky Way.}
\label{mwpressure}
\end{center}
\end{figure}

\subsection{Spectral Index Dependence on Hydrostatic Pressure:  A Leaky Box Model}
For our model to work requires a weak dependence of the CR electron
density, $N_{0}$, on HS pressure ($\S$ 5.3). The ordinary minimum 
energy argument instead accords equipartition to
the energy of the relativisic electrons
and that of the magnetic field, so that they both should
scale with $\bar P$. This argument would lead to a RC-CO correlation 
much steeper than observed: $I_{\rm RC} \propto I_{\rm CO}^{\,2.4}$,
whereas we instead observe an exponent that is lower by an order of magnitude.
Furthermore, it is commonly accepted that CR electrons diffuse from 
their injection sites. Thus, the CR electron density at a given 
location in the disk is the result of the balance between different 
processes such as the CR injection rate, the electron confinement time 
and energy losses.

To first approximation one might expect that the CR injection rate
should be proportional to the SFR surface density, and thus, to the FIR emission. 
The FIR-CO correlation (Figure~\ref{figuraflux}) would then lead to a 
 direct scaling of $N_{0}$ with $I_{\rm CO}$.
However, this in contrast with the hydrostatic pressure model which predicts a weak 
dependence of $N_{0}$ on $I_{\rm CO}$ (see previous section).

Let us consider the simple hypothesis that a galaxy, 
or any star-forming
region it contains, behaves essentially like a leaky box which is continuously 
replenished by a constant injection of fresh particles with a power law 
energy distribution at a rate 
$\partial N(\epsilon)/ \partial t=A\epsilon ^{-\gamma}$.
There are two time-scales which are relevant for the CR electron density. 
One is the confinement time of the electrons in the star-forming region, 
$t_{c}$; the other is the radiative life-time 
of the synchrotron electrons, $t_{\rm syn}$.  If $t_{c}\ll
t_{\rm syn}$, CR electrons will leave the galaxy disk (or the star-forming 
region) before losing their energy. 
That is, we expect that as the hydrostatic pressure increases, 
the star-formation rate grows, and with it the FIR emission and the 
cosmic ray injection rate.  However, the strong stellar winds associated 
with intense star-formation rates will facilitate the propagation of the ``excess'' cosmic rays 
from the disk to the halo.  In this regime, the synchrotron emission spectrum will be given
by Eq.~\ref{syncflux} with $N_{0}=A\,t_{c}$.
Under the assumption of hydrostatic pressure regulation, we found
that $N_{0}$ is nearly independent of $\bar P$ (${\S}$ 5.3); thus we
expect $t_{c} \propto 1/A$, and we expect that the observed
spectral index is about the same as the spectral index of the injected
particles.

\begin{figure}
\begin{center}
\includegraphics[width=8cm]{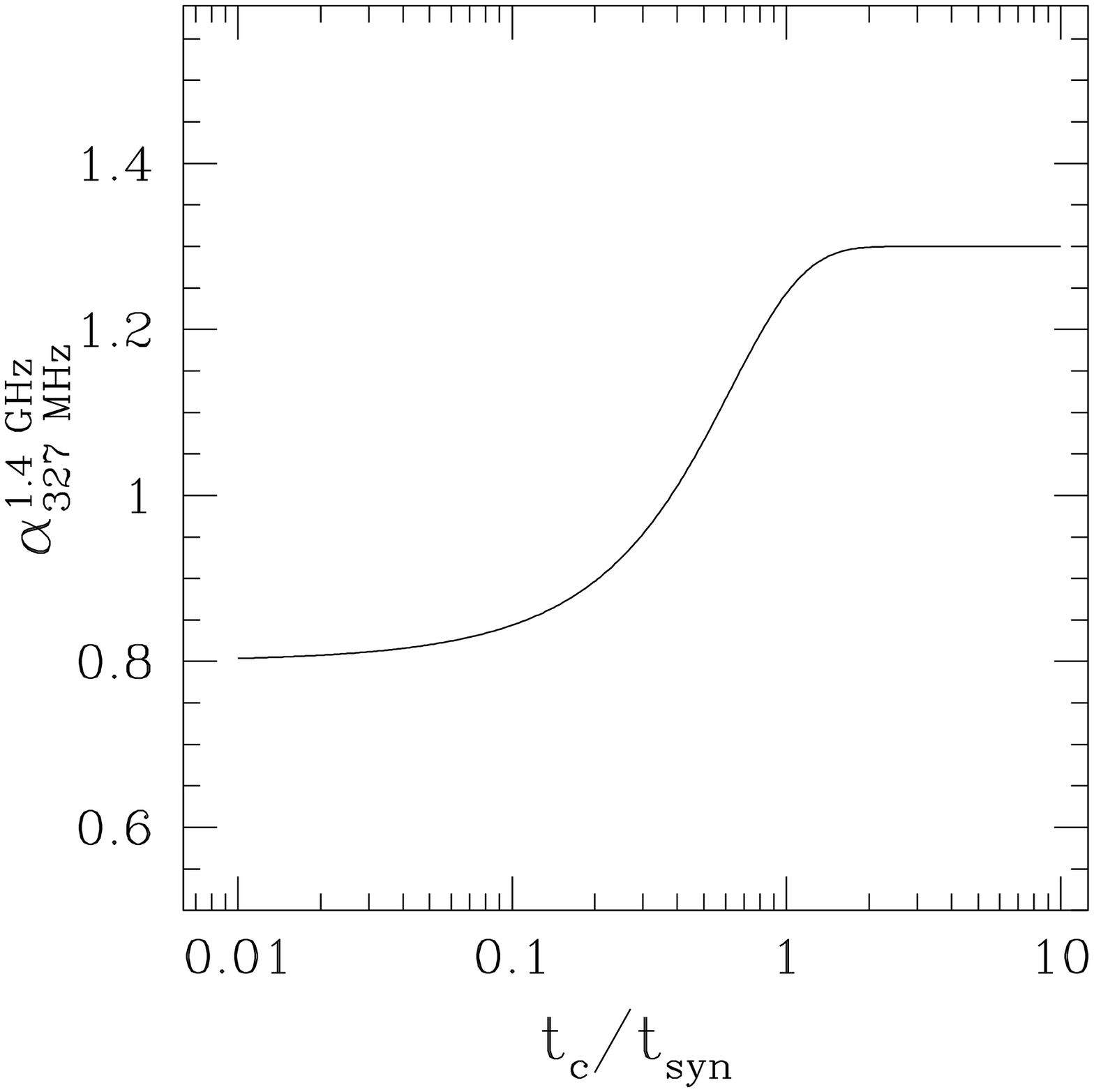}
\caption{Spectral index between 327 MHz and 1.4 GHz as a
  function of $t_{c}/t_{\rm syn}$ for a leaky-box model
  with an injection spectral index $\alpha_{0}=0.8$, see text.}
\label{alphatctsyn}
\end{center}
\end{figure}

As the hydrostatic pressure decreases, $t_{c}$ will progressively
increase.  When $t_{c}\gg t_{\rm syn}$, the CR electrons will lose all
their energy inside the parent star-forming region and the synchrotron
spectrum will be given by

\begin{equation}
I_{\rm RC} \propto N_0 B^{(\gamma +2)/2} \nu^{-\gamma/2}\;,
\label{syncflux2}
\end{equation}

\noindent with $N_{0}=A/B^{2}$. In this asymptotic regime, the spectral 
index of the synchrotron radiation emission is 0.5 steeper with respect to 
the injection spectral index; i.e. $I_{\rm RC}\propto \nu^{-1.3}$ for
an injection spectral index of $\alpha_{0}=0.8$ (see
Fig.~\ref{alphatctsyn}).

The electron synchrotron lifetime\footnote{The characteristic
  radiative age for electron emitting at 1\,GHz in a 10\,$\mu G$ magnetic field is
  $\sim3\times 10^{7}$ yr.} at a given frequency
scales as $t_{\rm syn} \propto B^{-3/2} \nu^{-1/2}$. Thus, under the 
hypothesis of energy
equipartition $B^2 \propto \bar P$, we have $t_{\rm syn} \propto
  \bar P^{-3/4}$.   In regions of strong star formation, where
$t_{c}\ll t_{\rm syn}$, we have seen that $I_{\rm RC} \propto
N_0 \bar{P}^{0.9}$ and $N_{0}=A\,t_{c}$; therefore
$I_{\rm RC} \propto A\,t_{c} \bar{P}^{0.9}$ and

\begin{equation}
\frac{t_{c}}{ t_{\rm syn}} \propto \frac{I_{\rm RC}\bar P^{\,-0.9}}{A
\bar P^{-3/4}}.
\label{tctsyn}
\end{equation}

\noindent
If we can further relate the particle injection rate $A$ to the
star formation rate (and more tentatively, to the FIR luminosity)
as $A = \Sigma_{SFR} = I_{\rm FIR}$, then

\begin{equation}
\frac{t_{c}}{ t_{\rm syn}} \propto \frac{I_{\rm RC}} 
{I_{\rm FIR}} \bar P^{-0.15}.
\label{tctsynfir}
\end{equation}

\begin{figure*}
\begin{center}
\includegraphics[width=14cm]{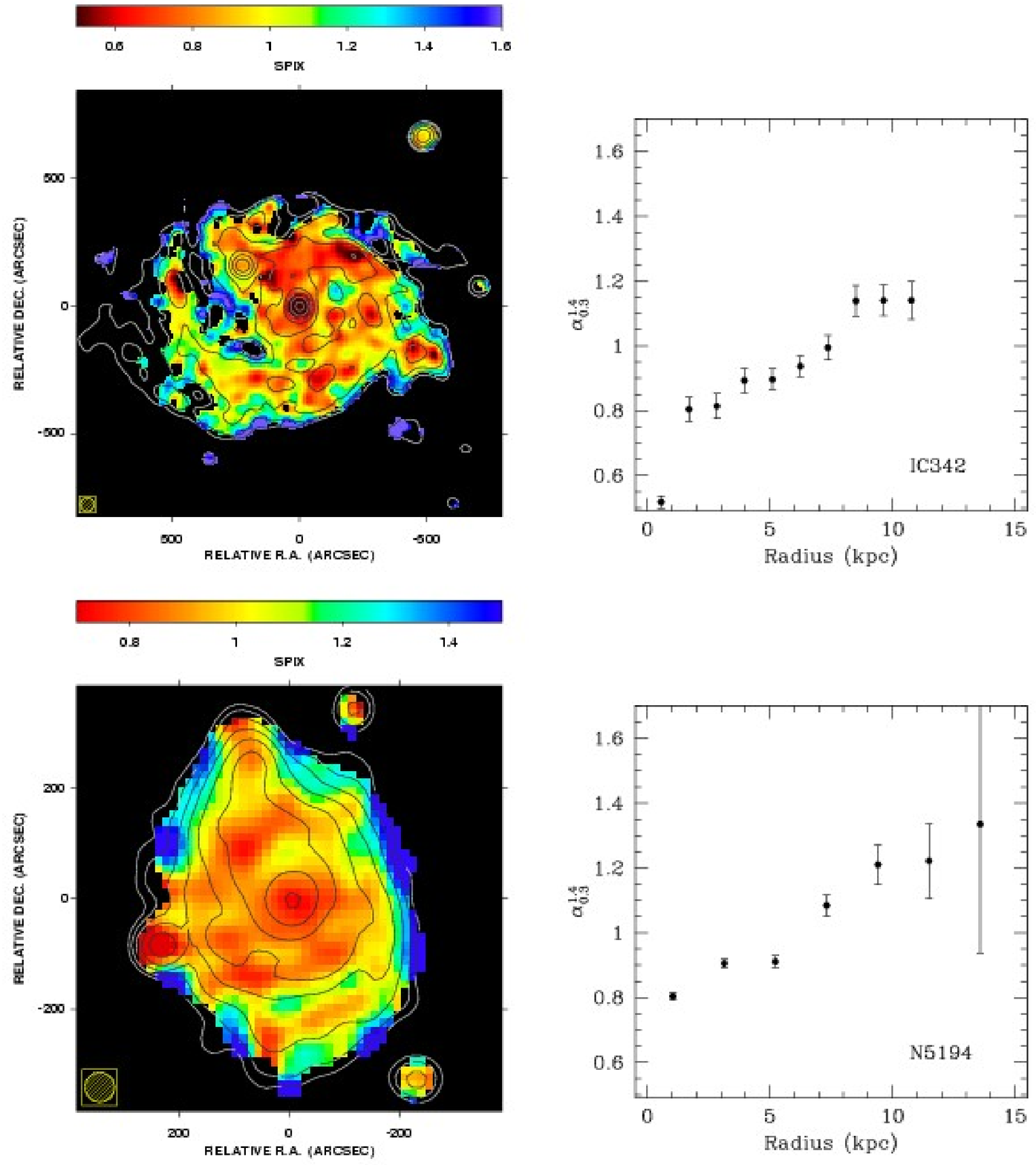}
\caption{Spectral index images (left panels) and radial averaged
  profiles (right panel) for IC 342 and NGC 5194. The spectral index
  is measured between 327 MHz and 1.4 GHz. Only pixels with an
  intensity level above $3\sigma$ at both frequencies have been considered.}
\label{spix}
\end{center}
\end{figure*}

We have seen in ${\S}$~4.4 that the FIR emission may have a shorter scale
length than the RC and CO emission, at least for galaxies
with bright CO and relatively strong star formation.  
According to Eq.~\ref{tctsynfir} and Fig.~\ref{alphatctsyn},
we expect that an increase of the RC/FIR ratio with radius will 
result in an average radial steepening of the
 synchrotron spectrum in galaxy disks.
In Fig.~\ref{spix} we present the spectral index images of IC~342 and 
NGC~5194, measured between 327 MHz and 1.4 GHz. 
The 327 MHz images are taken from the 
Westerbork Northern Sky Survey and their resolution, about 54\arcsec, is
well matched to that of the VLA D-array at 1.4 GHz. At these two
frequencies, the contribution from thermal emission is negligible.
Figure~\ref{spix} shows that the spiral arms are characterized by $\alpha
\simeq 0.8$. This indicates that these intense star-forming regions 
 suffer a significant CR electron leakage. According with the
 predictions of the hydrostatic pressure model, 
 in these high-pressure regions most probably $t_{c}\ll t_{\rm syn}$.  
The interarm regions, or in general the underlying disks, are
characterized by steeper spectra. The radially averaged spectral index
 profile show a monotonic steepening of the synchrotron spectrum with
 radius up to $\alpha \simeq 1.3-1.4$. This spectral behaviour 
suggests that as $\bar P$ decreases,
 the confinement time of the radiating electrons in the disk
 became larger than $t_{\rm syn}$.

It is worth noting that Niklas \& Beck (1997) used a similar argument
to test the phenomenological model of Helou \& Bicay (1993).
Niklas \& Beck concluded that on global scales, this model
failed to produce the expected dependence on spectral index
for galaxies with different radiation fields and magnetic field
strengths.
Indeed, the linearity of the global RC-FIR correlation, $S_{\rm RC} \propto
S_{\rm FIR} $ indicates that $\langle t_{c}/ t_{\rm
  syn} \rangle$ is, on
  average, constant. This explains why a relation between spectral
  index and magnetic field is not seen on global scales. 
However, we have extended this idea to local scales, within galaxy disks. 
According to Figs.~\ref{alphatctsyn} and \ref{spix}, we expect 
$t_{c}$ to be some fraction of $t_{\rm syn}$ within disks.

To summarize, we have used observationally determined relationships
between the molecular gas surface density and the
hydrostatic pressure to express the CO and RC intensities in terms of
$\bar{P}$. Both quantities scale with pressure since high-pressure
regions tend to be more molecular and, under the assumption of energy 
equipartition between the turbulent energy of the gas and the magnetic 
field energy, their synchrotron radiation amplified. 
For the hydrostatic pressure model to work requires 
a weak dependence of the cosmic ray electron density on pressure 
in disk galaxies. The observed radial gradient of cosmic ray
electrons in the Milky Way appears to support this hypothesis. If
cosmic ray leakage is taken into account, we show that this condition 
is compatible with a scenario in which the cosmic ray confinement time 
of the synchrotron electrons is inversely proportional to the SFR. 
The spatially resolved spectral index images of IC 342 and NGC 5194
are in general agreement with this prediction.

\section{Conclusions}

We have presented a comparison of the global RC, FIR, and CO
in 24 BIMA SONG galaxies and a detailed study of the high-resolution
(6\arcsec) CO-RC correlation in a subset of 9 sources.  Our main
conclusions are as follows:

(1) The global correlations between CO-RC, FIR-RC, and FIR-CO are
comparably tight, and we cannot identify any of the relations
as being a fundamental one.

(2) At 6\arcsec\ resolution, for CO-bright galaxies, 
the CO-RC correlation is as tightly
correlated as its global value, down to $\sim$ 100 pc size scales.
The correlation between the CO and RC emission is linear with
a dispersion that is less than a factor of two.

(3) Despite the low dispersion in the correlation, there are
systematic variations in \qco.  
The organized structures in \qco\ are found along spiral arms 
and on size scales much larger 
than the resolution of the observations.

(4) Combining the above two points suggests that, down to
100 pc scales, there does not appear to be a characteristic size 
scale where the CO-RC correlation breaks down.  

(5) Unlike the FIR-RC correlation, there is no systematic trend in
CO-RC as a function of radius in galaxies.  We speculate that 
high-resolution {\it Spitzer} observations
will show detailed spatial correlation between the FIR and RC and
CO.  We also expect that {\it Spitzer} will show that the dominant
variation in the FIR-RC correlation is across spiral arms, rather than in
a radial trend.

(6) We propose that the correlations between the RC and CO on 
large and small scales may be explained as a consequence of 
regulation by hydrostatic pressure in galaxies.  We use observationally
determined relationships between the molecular
gas surface density and the hydrostatic pressure to
derive a relationship between the CO and RC intensities.
Taken together with recent results linking the FIR emission with
ISM pressure by Dopita et al. (2004), these scenarios
link all three of the CO-RC-FIR emissions with the 
interstellar pressure, without invoking any explicit dependence
on a star formation scenario.

(7) Under the assumption of hydrostatic pressure regulation, 
we use a leaky box model of cosmic rays to show that the
spectral index between 327 MHz and 1.4 GHz depends on the
ratio of the confinement time of the electrons in the
star-forming region, $t_{c}$, to their synchrotron lifetime,
$t_{\rm syn}$.  We expect that in regions of high hydrostatic
pressure, $t_{c} \ll t_{\rm syn}$ and $\alpha \approx 0.8$.
As the pressure decreases, the electrons decay before
escaping the disk, $t_{c} \gg t_{\rm syn}$, and the spectral 
index steepens to $\alpha \approx 1.3$.
We show spectral index maps of IC~342 and NGC~5194 that
reflect this kind of dependence, with $\alpha \approx\ 0.8$
in the spiral arms and a monotonic steepening with
radius up to $\alpha \approx\ 1.3 - 1.4$.

\begin{acknowledgements}
The authors thank an anonymous referee for the valuable comments and 
suggestions. We thank Hans Bloemen, Alberto Bolatto, Al Glassgold, Anne Green, 
Wim Hermsen, David Hollenbach, John Lugten, Chris McKee, Andy Strong, 
and Peter Teuben for helpful discussions. We are grateful to 
Rainer Beck, who provided us with his M51 radio-continuum image.
Erik Rosolowsky kindly produced Figure \ref{mwpressure} for us.
T.T.H. is grateful to M. W. Regan for 
a helpful exchange regarding the CO scale lengths.  
This work is partially supported by NSF grant 0228963 to the Radio 
Astronomy Laboratory at UC Berkeley. M.M. wish to thank the  
Berkeley Radio Astronomy Laboratory for their hospitality 
during the development of this work.
\end{acknowledgements}

\appendix

\section{Notes on Individual Galaxies}

{\bf IC 342}. At d=3.9 Mpc, IC 342 is the nearest galaxy in the 
sample, and the high-resolution images provide linear resolutions
of $\sim$100 pc.  (A recent Cepheid distance determination by
Saha, Claver, \& Hoessel 2002 suggests that d=3.3 Mpc, which
would provide slightly improved linear resolution.)
A prominent (though kinematically weak) north-south bar dominates 
the CO emission in the inner 2\arcmin\ of this starburst source, 
with two spiral arms at 
larger distances north and south of the bar (Crosthwaite et al. 2001).
Within the central bar, the scatter of the correlation 
is higher than it is at larger distances.  The most 
significant deviation from the correlation is found in the inner 15\arcsec~
where the peak of the RC emission is $\sim10$ times higher with respect 
to the CO emission.  This starburst galaxy may be a large-scale
analog to starforming regions in the Milky Way, which do not exhibit
the RC-FIR correlation on size scales of a few hundred pc (e.g.
Boulanger \& Perault 1988).
However, it is relevant to note that on larger (55\arcsec, or 1.0 kpc)
scales, the CO-RC correlation remains tight.

{\bf NGC 4258}.  The radio emission in this unusual source is associated 
with two, large-scale ``anomalous arms'' not seen in the optical but detected
first in H$\alpha$ (Courtes \& Cruvellier 1961).  In the inner
galaxy, the radio emission is clearly associated with jets,
presumably related to the AGN activity associated with the
supermassive black hole; however, at larger distances,
the nature of the radio emission is not so clear (e.g. 
Krause \& L{\" o}hr 2004 and references therein).   A detailed
discussion of this unusual galaxy is beyond the scope of this
paper, and a proper discussion of the RC-FIR-CO correlations in
this source would warrant a separate study.  However, we include
NGC 4258 here and emphasize that, even in this highly unusual 
source, the CO-RC correlation is very good over those regions
where both CO and RC are present.

{\bf NGC 4414}.  The ``low-resolution'' RC emission from this source
has been convolved from 14\arcsec\ resolution up to 55\arcsec,
and at 55\arcsec, the RC emission as well as the CO emission are only 
marginally resolved.  One must therefore be somewhat cautious
about whether the 55\arcsec\ pixels shown in the point-by-point
comparison in the upper right panel of Figure 1($c$) are truly
independent.  However, at 6\arcsec, the CO and RC structures of
this galaxy are revealed as a ring with a diameter of $\sim$ 5 kpc.
NGC 4414 is a flocculent galaxy, which means that it lacks prominent
spiral structure in optical light, though Thornley (1996) found 
that $K\arcmin$ imaging revealed low-level spiral structure. 
However, of four sources with newly-revealed spiral arms
presented by Thornley, NGC 4414 remained ``the most
flocculent of the sample in $K\arcmin$ emission''.
The \qco\ image shown in Figure \ref{figuraq} appears to be
nearly featureless and shows a dispersion of only 
0.09; that is, the CO/RC intensity ratio varies by only a
factor of 1.2.
On the other hand, the dynamic range of emission in the CO and
RC varies by only slightly more than one order of magnitude.

{\bf NGC 4736}.
The molecular gas in this nearby source is distributed in a central 
molecular bar and in two tightly wound spiral arms which are associated 
with a bright ring of star formation (Wong \& Blitz 2000).  A background
radio continuum source, apparent in the upper left panel of Figure
1($d$), has been omitted from the analysis in this paper.   As discussed
in ${\S}$4.3, there appears to be an enhancement in \qco\ 
(Figure \ref{figuraq}) associated with the northwestern spiral
arm; the enhancement may be a result of the starburst there.
It also appears that \qco\ is larger at the bar ends,
at roughly $\pm$15\arcsec\ from the galaxy center, and along the
``arclike'' extensions from the bar that probably are associated
with the beginnings of the spiral structure (Wong \& Blitz 2000).

{\bf NGC 5005}.
Two background continuum sources are apparent in the low-resolution
RC images of NGC 5005; these have been omitted from all analysis.
Like NGC 4414, the 55\arcsec\ observations show that the CO and
RC is only slightly resolved.  At 6\arcsec\ resolution, the
central unresolved RC emission is associated with a nonstellar
active galactic nucleus (AGN).  Outside the nucleus, the agreement
between the RC and CO emission is excellent, as can be seen
both in Figure 1($e$) as well as in Figure \ref{figuraq};
the \qco\ image is nearly featureless and has low dispersion.

{\bf NGC 5033}.
Like NGC 5005, NGC 5033 has an unresolved AGN at its nucleus,
and its \qco\ image is nearly featureless.  An enhancement in
\qco\ is seen at the southern end of the CO emission;
however, this CO emission is detected only at a 
$\sim$ 3$\sigma$ level and might be spurious.
Background continuum sources are apparent in both the high-
and low-resolution data sets.

{\bf NGC 5055}.  NGC 5055 is the prototypical flocculent spiral
galaxy, which (like NGC 4414) was revealed to have regular
spiral structure in $K\arcmin$-band imaging (Thornley 1996).
At 6\arcsec\ resolution, the RC and the CO emissions appear nearly
featureless, though an inspection of the CO image and especially the
\qco\ image (Figure \ref{figuraq}) show structures that follow
very closely what appear to be inner spiral arms in Thornley's
$K\arcmin$-band image.   Nonetheless, the dispersion in
\qco\ is only 0.11.

{\bf NGC 5194}.
The Whirlpool galaxy M~51 is the prototypical grand design
spiral galaxy, with two prominent spiral arms where the
molecular gas is generally confined.  The arms wind in
towards the center of the galaxy but remain in two
distinct peaks; the molecular gas lacks emission right at
the nucleus.  The RC at the nucleus is dominated by
the AGN activity.  The \qco\ image is remarkable in 
that it shows strong enhancements along the spiral arms,
as discussed in ${\S}$4.3.

{\bf NGC 6946}.
This nearby starburst galaxy has an elongated, bright 
north-south molecular feature in the inner $\sim$ 50\arcsec\ and
wispy arms at larger radii.  The \qco\ image (Figure \ref{figuraq})
and profile (Figure 1$i$) show an enhanced ratio within
the central region, apart from the compact HII region nucleus, which
shows relatively enhanced RC emission.  Like IC 342, 
it may be that this region is a large-scale analog to
active starforming regions in the Milky Way.

\end{document}